\documentclass[%
 reprint,
nofootinbib,
 amsmath,amssymb,
 aps,
]{revtex4-2}

\usepackage{graphicx}
\usepackage{dcolumn}
\usepackage{bm}
\usepackage{upgreek}
\usepackage[mathlines]{lineno}
\usepackage{enumitem}
\usepackage{orcidlink}

\newcommand{\dif}{\mathrm{d}}

\newcommand{\LL}{\mathrm{L}}
\newcommand{\RR}{\mathrm{R}}
\newcommand{\st}{\mathrm{s}}
\newcommand{\elec}{\mathrm{e}}

\begin{document}


\title{A search for sterile neutrinos at ultrahigh energies at the Pierre Auger Observatory}%

\author{%
A.~Abdul Halim\orcidlink{0000-0003-1218-1737}$^{13}$,
P.~Abreu\orcidlink{0000-0002-9973-7314}$^{67}$,
M.~Aglietta\orcidlink{0000-0001-8354-5388}$^{50,49}$,
M.~Ahmed\orcidlink{0009-0007-7727-5052}$^{31}$,
I.~Allekotte\orcidlink{0000-0002-9624-3044}$^{1}$,
K.~Almeida Cheminant\orcidlink{0000-0001-6352-5339}$^{74,73}$,
R.~Aloisio\orcidlink{0000-0003-0161-5923}$^{42,43}$,
J.~Alvarez-Mu\~niz\orcidlink{0000-0002-2367-0803}$^{72}$,
A.~Ambrosone\orcidlink{0000-0002-9942-1029}$^{42,43}$,
J.~Ammerman Yebra\orcidlink{0000-0002-2059-6206}$^{73}$
% ...

}

\affiliation{%
\begin{description}[labelsep=0.2em,align=right,labelwidth=0.7em,labelindent=0em,leftmargin=2em,noitemsep,before={\renewcommand\makelabel[1]{##1 }}]
\item[$^{1}$] Centro At\'omico Bariloche and Instituto Balseiro (CNEA-UNCuyo-CONICET), San Carlos de Bariloche, Argentina
\item[$^{2}$] Departamento de F\'\i{}sica and Departamento de Ciencias de la Atm\'osfera y los Oc\'eanos, FCEyN, Universidad de Buenos Aires and CONICET, Buenos Aires, Argentina
\item[$^{3}$] IFLP, Universidad Nacional de La Plata and CONICET, La Plata, Argentina
\item[$^{4}$] Instituto de Astronom\'\i{}a y F\'\i{}sica del Espacio (IAFE, CONICET-UBA), Buenos Aires, Argentina
\item[$^{5}$] Instituto de F\'\i{}sica de Rosario (IFIR) -- CONICET/U.N.R.\ and Facultad de Ciencias Bioqu\'\i{}micas y Farmac\'euticas U.N.R., Rosario, Argentina
\item[$^{6}$] Instituto de Tecnolog\'\i{}as en Detecci\'on y Astropart\'\i{}culas (CNEA, CONICET, UNSAM), and Universidad Tecnol\'ogica Nacional -- Facultad Regional Mendoza (CONICET/CNEA), Mendoza, Argentina
\item[$^{7}$] Instituto de Tecnolog\'\i{}as en Detecci\'on y Astropart\'\i{}culas (CNEA, CONICET, UNSAM), Buenos Aires, Argentina
\item[$^{8}$] International Center of Advanced Studies and Instituto de Ciencias F\'\i{}sicas, ECyT-UNSAM and CONICET, Campus Miguelete -- San Mart\'\i{}n, Buenos Aires, Argentina
\item[$^{9}$] Laboratorio Atm\'osfera -- Departamento de Investigaciones en L\'aseres y sus Aplicaciones -- UNIDEF (CITEDEF-CONICET), Argentina
\item[$^{10}$] Observatorio Pierre Auger, Malarg\"ue, Argentina
\item[$^{11}$] Observatorio Pierre Auger and Comisi\'on Nacional de Energ\'\i{}a At\'omica, Malarg\"ue, Argentina
\item[$^{12}$] Universidad Tecnol\'ogica Nacional -- Facultad Regional Buenos Aires, Buenos Aires, Argentina
\item[$^{13}$] Adelaide University, Adelaide, S.A., Australia
\item[$^{14}$] Universit\'e Libre de Bruxelles (ULB), Brussels, Belgium
\item[$^{15}$] Vrije Universiteit Brussels, Brussels, Belgium
\item[$^{16}$] Centro Brasileiro de Pesquisas Fisicas, Rio de Janeiro, RJ, Brazil
\item[$^{17}$] Centro Federal de Educa\c{c}\~ao Tecnol\'ogica Celso Suckow da Fonseca, Petropolis, Brazil
\item[$^{18}$] Universidade de S\~ao Paulo, Escola de Engenharia de Lorena, Lorena, SP, Brazil
\item[$^{19}$] Universidade de S\~ao Paulo, Instituto de F\'\i{}sica de S\~ao Carlos, S\~ao Carlos, SP, Brazil
\item[$^{20}$] Universidade de S\~ao Paulo, Instituto de F\'\i{}sica, S\~ao Paulo, SP, Brazil
\item[$^{21}$] Universidade Estadual de Campinas (UNICAMP), IFGW, Campinas, SP, Brazil
\item[$^{22}$] Universidade Estadual de Feira de Santana, Feira de Santana, Brazil
\item[$^{23}$] Universidade Federal do ABC, Santo Andr\'e, SP, Brazil
\item[$^{24}$] Universidade Federal do Paran\'a, Setor Palotina, Palotina, Brazil
\item[$^{25}$] Universidade Federal do Rio de Janeiro, Instituto de F\'\i{}sica, Rio de Janeiro, RJ, Brazil
\item[$^{26}$] Universidad de Medell\'\i{}n, Medell\'\i{}n, Colombia
\item[$^{27}$] Universidad Industrial de Santander, Bucaramanga, Colombia
\item[$^{28}$] Charles University, Faculty of Mathematics and Physics, Institute of Particle and Nuclear Physics, Prague, Czech Republic
\item[$^{29}$] Institute of Physics of the Czech Academy of Sciences, Prague, Czech Republic
\item[$^{30}$] Palacky University, Olomouc, Czech Republic
\item[$^{31}$] CNRS/IN2P3, IJCLab, Universit\'e Paris-Saclay, Orsay, France
\item[$^{32}$] Laboratoire de Physique Nucl\'eaire et de Hautes Energies (LPNHE), Sorbonne Universit\'e, Universit\'e de Paris, CNRS-IN2P3, Paris, France
\item[$^{33}$] Universit\'e Paris-Saclay, CNRS/IN2P3, IJCLab, Orsay, France
\item[$^{34}$] Bergische Universit\"at Wuppertal, Department of Physics, Wuppertal, Germany
\item[$^{35}$] Karlsruhe Institute of Technology (KIT), Institute for Experimental Particle Physics, Karlsruhe, Germany
\item[$^{36}$] Karlsruhe Institute of Technology (KIT), Institut f\"ur Prozessdatenverarbeitung und Elektronik, Karlsruhe, Germany
\item[$^{37}$] Karlsruhe Institute of Technology (KIT), Institute for Astroparticle Physics, Karlsruhe, Germany
\item[$^{38}$] RWTH Aachen University, III.\ Physikalisches Institut A, Aachen, Germany
\item[$^{39}$] TU Dortmund University, Department of Physics, Dortmund, Germany
\item[$^{40}$] Universit\"at Hamburg, II.\ Institut f\"ur Theoretische Physik, Hamburg, Germany
\item[$^{41}$] Universit\"at Siegen, Department Physik -- Experimentelle Teilchenphysik, Siegen, Germany
\item[$^{42}$] Gran Sasso Science Institute, L'Aquila, Italy
\item[$^{43}$] INFN Laboratori Nazionali del Gran Sasso, Assergi (L'Aquila), Italy
\item[$^{44}$] INFN, Sezione di Catania, Catania, Italy
\item[$^{45}$] INFN, Sezione di Lecce, Lecce, Italy
\item[$^{46}$] INFN, Sezione di Milano, Milano, Italy
\item[$^{47}$] INFN, Sezione di Napoli, Napoli, Italy
\item[$^{48}$] INFN, Sezione di Roma ``Tor Vergata'', Roma, Italy
\item[$^{49}$] INFN, Sezione di Torino, Torino, Italy
\item[$^{50}$] Osservatorio Astrofisico di Torino (INAF), Torino, Italy
\item[$^{51}$] Politecnico di Milano, Dipartimento di Scienze e Tecnologie Aerospaziali , Milano, Italy
\item[$^{52}$] Universit\`a del Salento, Dipartimento di Matematica e Fisica ``E.\ De Giorgi'', Lecce, Italy
\item[$^{53}$] Universit\`a dell'Aquila, Dipartimento di Scienze Fisiche e Chimiche, L'Aquila, Italy
\item[$^{54}$] Universit\`a di Catania, Dipartimento di Fisica e Astronomia ``Ettore Majorana``, Catania, Italy
\item[$^{55}$] Universit\`a di Milano, Dipartimento di Fisica, Milano, Italy
\item[$^{56}$] Universit\`a di Napoli ``Federico II'', Dipartimento di Fisica ``Ettore Pancini'', Napoli, Italy
\item[$^{57}$] Universit\`a di Palermo, Dipartimento di Fisica e Chimica ''E.\ Segr\`e'', Palermo, Italy
\item[$^{58}$] Universit\`a di Roma ``Tor Vergata'', Dipartimento di Fisica, Roma, Italy
\item[$^{59}$] Universit\`a Torino, Dipartimento di Fisica, Torino, Italy
\item[$^{60}$] Benem\'erita Universidad Aut\'onoma de Puebla, Puebla, M\'exico
\item[$^{61}$] Unidad Profesional Interdisciplinaria en Ingenier\'\i{}a y Tecnolog\'\i{}as Avanzadas del Instituto Polit\'ecnico Nacional (UPIITA-IPN), M\'exico, D.F., M\'exico
\item[$^{62}$] Universidad Aut\'onoma de Chiapas, Tuxtla Guti\'errez, Chiapas, M\'exico
\item[$^{63}$] Universidad Michoacana de San Nicol\'as de Hidalgo, Morelia, Michoac\'an, M\'exico
\item[$^{64}$] Universidad Nacional Aut\'onoma de M\'exico, M\'exico, D.F., M\'exico
\item[$^{65}$] Institute of Nuclear Physics PAN, Krakow, Poland
\item[$^{66}$] University of \L{}\'od\'z, Faculty of High-Energy Astrophysics,\L{}\'od\'z, Poland
\item[$^{67}$] Laborat\'orio de Instrumenta\c{c}\~ao e F\'\i{}sica Experimental de Part\'\i{}culas -- LIP and Instituto Superior T\'ecnico -- IST, Universidade de Lisboa -- UL, Lisboa, Portugal
\item[$^{68}$] ``Horia Hulubei'' National Institute for Physics and Nuclear Engineering, Bucharest-Magurele, Romania
\item[$^{69}$] Institute of Space Science, Bucharest-Magurele, Romania
\item[$^{70}$] Center for Astrophysics and Cosmology (CAC), University of Nova Gorica, Nova Gorica, Slovenia
\item[$^{71}$] Experimental Particle Physics Department, J.\ Stefan Institute, Ljubljana, Slovenia
\item[$^{72}$] Instituto Galego de F\'\i{}sica de Altas Enerx\'\i{}as (IGFAE), Universidade de Santiago de Compostela, Santiago de Compostela, Spain
\item[$^{73}$] IMAPP, Radboud University Nijmegen, Nijmegen, The Netherlands
\item[$^{74}$] Nationaal Instituut voor Kernfysica en Hoge Energie Fysica (NIKHEF), Science Park, Amsterdam, The Netherlands
\item[$^{75}$] Stichting Astronomisch Onderzoek in Nederland (ASTRON), Dwingeloo, The Netherlands
\item[$^{76}$] Case Western Reserve University, Cleveland, OH, USA
\item[$^{77}$] Colorado School of Mines, Golden, CO, USA
\item[$^{78}$] Department of Physics and Astronomy, Lehman College, City University of New York, Bronx, NY, USA
\item[$^{79}$] Michigan Technological University, Houghton, MI, USA
\item[$^{80}$] New York University, New York, NY, USA
\item[$^{81}$] University of Chicago, Enrico Fermi Institute, Chicago, IL, USA
\item[$^{82}$] University of Delaware, Department of Physics and Astronomy, Bartol Research Institute, Newark, DE, USA
\item[] -----
\item[$^{a}$] Max-Planck-Institut f\"ur Radioastronomie, Bonn, Germany
\item[$^{b}$] also at Kapteyn Institute, University of Groningen, Groningen, The Netherlands
\item[$^{c}$] School of Physics and Astronomy, University of Leeds, Leeds, United Kingdom
\item[$^{d}$] Fermi National Accelerator Laboratory, Fermilab, Batavia, IL, USA (Affiliation for identification purposes only)
\item[$^{e}$] Pennsylvania State University, University Park, PA, USA
\item[$^{f}$] Colorado State University, Fort Collins, CO, USA
\item[$^{g}$] Louisiana State University, Baton Rouge, LA, USA
\item[$^{h}$] Universit\'e Grenoble Alpes, CNRS, Grenoble Institute of Engineering, LPSC-IN2P3, Grenoble, France
\item[$^{i}$] Institut universitaire de France (IUF), France
\item[$^{j}$] now at Graduate School of Science, Osaka Metropolitan University, Osaka, Japan
\item[$^{k}$] also at: Universit\`a degli Studi di Trento, Dipartimento di Fisica, Trento, Italy
\end{description}

}

\collaboration{The Pierre Auger Collaboration}
\email{spokespersons@auger.org}
\homepage{http://www.auger.org}
\noaffiliation

\date{\today}

\begin{abstract}

We investigate the sensitivity of the Pierre Auger Observatory to sterile neutrinos with masses in the MeV-to-100\,MeV range. To this end, we consider a minimal $3+1$ framework in which a single  sterile state mixes mainly with the $\nu_\uptau$ flavor through an active--sterile mixing angle $\theta_N$ to produce a mass eigenstate $N_1$. At ultrahigh energies, $N_1$ neutrinos can traverse substantial chord lengths through the Earth and interact through their active component, producing $\uptau$ leptons that may emerge near the surface and decay in the atmosphere, thereby generating detectable air showers. We compute the corresponding exposures with a detailed Monte Carlo simulation of neutrino propagation, interactions, $\uptau$ energy losses and decay, and derive upper limits at $90\%$ confidence level on an isotropic diffuse  flux of $N_1$ neutrinos. For an $E^{-2}$ spectrum, we obtain $E^2\Phi_N\leq 2\times10^{-9}\sin^{-2}\theta_N\,\mathrm{GeV\,cm^{-2}\,sr^{-1}\,s^{-1}}$ over $0.3$--$200$ EeV for $\sin^2\theta_N\lesssim10^{-1}$. We further consider $N_1$ neutrinos arising from the decay of a superheavy dark-matter particle. For a benchmark dark-matter mass of $10^{13}$ GeV, we derive constraints on the active--sterile mixing angle and dark-matter lifetime for sterile-neutrino masses of $10$ and $100$ MeV. These results illustrate the potential of the Observatory to probe long-lived weakly interacting particles.

\end{abstract}

\maketitle


\section{\label{sec:intro}Introduction}

The observation of neutrino oscillations establishes that they are massive and therefore provides evidence for physics beyond the Standard Model. Sterile states introduced in seesaw frameworks offer a compelling explanation for the origin of these masses~\cite{Gell-Mann:1979vob,Yanagida1979}. While the conventional seesaw mechanism relates the smallness of the light-neutrino masses to a high scale, low-scale realizations in which an approximate lepton-number symmetry suppresses the light-neutrino masses are also viable~\cite{Mohapatra:1986bd,Mohapatra:1986aw,Montero:2025hye,Anchordoqui:2026ret}. Such low-scale scenarios allow for sterile states with masses in the MeV--GeV range and with active--sterile mixings potentially accessible to experiment~\cite{Kersten:2007vk,Drewes:2016jae,Domcke:2020ety}. They are also of interest for leptogenesis, in which the dynamics of sterile states in the early Universe can generate the observed baryon asymmetry~\cite{Asaka:2005an,Drewes:2012ma,Drewes:2016jae}. These considerations motivate searches for sterile neutrinos from astrophysical origin.

The Pierre Auger Observatory~\cite{PierreAuger:2015eyc} is particularly well suited to probe weakly interacting particles such as sterile neutrinos that would traverse substantial fractions of the diameter of the Earth before producing secondary $\uptau$ leptons. In the ultrahigh-energy range considered in this work (approximately $10^{17}$--$10^{21}$ \,eV), $\uptau$ leptons experience radiative energy losses in matter that remain sufficiently moderate compared with those of lighter charged leptons. In this manner, $\uptau$ leptons produced near the surface can emerge from the Earth and subsequently decay in the atmosphere, generating extensive air showers detectable through the ``Earth-skimming channel'' by the Surface Detector of the Observatory~\cite{Bertou:2001vm,PierreAuger:2007vvh,PierreAuger:2019ens,PierreAuger:2023wke} or through the ``upward-going channel'' by its Fluorescence Detector~\cite{PierreAuger:2025hvl}. This detection strategy is therefore sensitive not only to Standard Model $\uptau$ neutrinos but also to a broad class of feebly interacting particles that regenerate $\uptau$ leptons near the surface of the Earth~\cite{Cherry:2018rxj,Huang:2018als,Anchordoqui:2018ucj,Anchordoqui:2018ssd,Chauhan:2018lnq,Collins:2018jpg,Heurtier:2019rkz,Heurtier:2019git,Hooper:2019ytr,Esteban:2019hcm,Chipman:2019vjm,Abdullah:2019ofw,Borah:2019ciw,Esmaili:2019pcy,Cline:2019snp,Altmannshofer:2020axr,Liang:2021rnv,Heighton:2023qpg,Bertolez-Martinez:2023scp,Anchordoqui:2025xug}.

The Earth-skimming channel, corresponding to zenith angles between $90^\circ$ and $95^\circ$, provides the largest exposure to $\uptau$ leptons produced in the Earth. In parallel, the upward-going channel, involving trajectories that traverse much larger chord lengths through the Earth, provides a valuable diagnostic tool: while Earth-skimming events alone cannot distinguish between expected Standard Model $\uptau$ neutrinos and beyond-Standard-Model particles that regenerate $\uptau$ leptons, the observation of upward-going events would provide a much more distinctive signature of new weakly interacting states.

In this work, we consider a $3+1$ framework in which one sterile state mixes appreciably only with the $\uptau$ flavor. The physical mass eigenstate $N_1$, predominantly sterile, interacts through its small active component, with interaction rates suppressed by the active--sterile mixing parameter. In the MeV-to-100\,MeV mass range considered here, matter effects on sterile--active propagation inside the Earth remain negligible throughout the ultrahigh-energy range of interest, such that this suppression is governed by the vacuum mixing angle. An ultrahigh-energy $N_1$ neutrino entering the Earth can consequently traverse a substantial chord before interacting through its active component and lead to an observable $\uptau$-induced shower. The 3+1 framework therefore provides a natural benchmark for demonstrating the potential of the upward-going channel to distinguish beyond-the-Standard-Model from the Standard Model signals.

Historically, upward-going air-shower searches received particular attention following two anomalous radio pulses reported by the ANITA-I~\cite{ANITA:2016vrp} and ANITA-III~\cite{ANITA:2018sgj} balloon flights, which appeared compatible with upward-going air showers emerging from the Earth at energies and emergence angles difficult to accommodate within Standard Model expectations~\cite{Halzen:1998be,Bigas:2008sw,Alvarez-Muniz:2018owm}. These events motivated dedicated searches for upward-going showers with zenith angles exceeding $110^\circ$ using the Fluorescence Detector of the Observatory~\cite{PierreAuger:2025hvl}. No significant excess over the expected background was observed.

Using detailed Monte Carlo simulations of neutrino propagation through the Earth, neutrino interactions, $\uptau$-lepton energy losses, and $\uptau$ decay, we calculate the exposure of the Observatory to the $N_1$-neutrino-induced air showers and derive upper limits on a diffuse flux of $N_1$ neutrinos. We investigate two possible sources of the ultrahigh-energy sterile neutrinos. First, we consider a diffuse cosmogenic flux generated by active--sterile oscillations during propagation from astrophysical sources to the Earth. The corresponding sterile component is obtained using phase-averaged flavor-transition probabilities in the $3+1$ framework and is subsequently attenuated by sterile-neutrino decay during cosmological propagation.  Second, we consider a nonminimal production mechanism in which sterile neutrinos are produced directly in the decay of a superheavy dark-matter particle. In this case, the sterile neutrinos can constitute the only portal to a dark sector that is otherwise decoupled from the Standard Model. We derive constraints on the active--sterile mixing angle and dark-matter lifetime for representative sterile-neutrino masses of $10$ and $100$ MeV. All resulting constraints are compared with, and placed in the context of, existing bounds from beam-dump experiments and cosmology.

\section{\label{sec:theory} Standard model extension with low-scale sterile neutrinos}

We consider a generic type-I seesaw extension of the Standard Model (SM) in which $n$ right-handed (sterile) neutrinos $N_{\st I}$ $(I=1,\dots,n)$ are added to the particle content. After electroweak symmetry breaking, the neutrino squared-mass matrix can be diagonalized to yield $3+n$ Majorana mass eigenstates. These consist of three predominantly active neutrinos $\nu_{i}$ $(i=1,2,3)$ and $n$ predominantly sterile states $N_{I}$ $(I=1,\dots,n)$. In our benchmark, only one sterile neutrino mixes appreciably with the active sector. In particular, we assume that this mixing occurs predominantly with the third lepton flavor, i.e., the $\uptau$-flavored neutrino, while the remaining sterile states are effectively decoupled from the low-energy phenomenology. In this sector, experimental limits on active--sterile mixing angles remain relatively weak due to the challenges inherent in tau neutrino production and detection. Under this assumption, the flavor eigenstate $\nu_{\uptau_\LL}$ is related to the mass eigenstates through an extended $4 \times 4$ unitary mixing matrix $U$,
\begin{equation}
    \nu_{\uptau\LL}=\sum_{i=1}^3U_{\uptau i}\nu_{i\LL}+U_{\uptau 4}(N_{1\RR})^\mathrm{c},
\end{equation}
where $(N_{1\RR})^\mathrm{c}$ is the field conjugate of $N_{1\RR}$ that satisfies $(N_{1\RR})^\mathrm{c}=(N_1^\mathrm{c})_\LL$. In this parameterization, the active--sterile mixing angle is defined from the oscillation amplitude as $\sin^2\theta_N=|U_{\uptau 4}|^2$.  For brevity, we refer to $N_1$ as the ``sterile neutrino'' throughout the following, with the understanding that it is a predominantly sterile mass eigenstate.

$N_1$ inherits suppressed weak interactions proportional to $\sin^2\theta_N$ via its mixing with the $\uptau$ active flavor. An additional consideration is the effect of matter on active--sterile neutrino propagation. As a quasi-sterile neutrino traverses the Earth, coherent forward scattering of the active component modifies the effective active--sterile mixing angle and can therefore alter the interaction probability relative to that expected from vacuum oscillations. For the predominantly $\nu_\uptau$ mixing considered here, the relevant matter potential $V_\mathrm{m}$ arises from neutral-current interactions with neutrons~\cite{Wolfenstein:1977ue,Dasgupta:2021ies}. In a two-state approximation, the effective mixing angle is
\begin{equation}
\sin^2 2\theta_N^{\rm m}
=
\frac{\Delta^2\sin^2 2\theta_N}
{\left(\Delta\cos2\theta_N-V_{\rm m}\right)^2+\Delta^2\sin^2 2\theta_N},
\end{equation}
where $V_{\rm m}\simeq-G_F N_n/\sqrt{2}$ and $\Delta=(m_N-m_\nu)^2/2E$, with $G_\mathrm{F}$ the Fermi constant,  $N_n$ the neutron density, $E$ the energy, and $m_N$ and $m_\nu$ the masses of the $N_1$ and light neutrino states. Matter-induced suppression becomes relevant when $2E|V_{\rm m}|\gtrsim (m_N-m_\nu)^2$, whereas the vacuum limit is recovered for $2E|V_{\rm m}|\ll  (m_N-m_\nu)^2$. For typical Earth densities, $|V_{\rm m}|\sim10^{-13}$ eV, such that even at $E=10^{21}$ eV the corresponding matter scale is $2E|V_{\rm m}|\sim10^8$ eV$^2$. This is much smaller than $( m_N-m_\nu)^2\simeq m_N^2\geq10^{12}$ eV$^2$ for $m_N\geq1$ MeV. Matter effects are therefore negligible throughout the parameter space considered here, and the sterile-neutrino interaction probability can be calculated using the vacuum mixing angle, $\theta_N$.

In the vanilla realization of the type-I seesaw mechanism, the Majorana mass scale of the sterile states is typically assumed to lie far above the electroweak scale, $m_\RR\gg v$. This  accounts for the smallness of the masses of the predominantly active states while giving rise to predominantly sterile states with masses $m_N\simeq m_\RR$. However, the seesaw mechanism does not fundamentally require the sterile-neutrino mass to lie at a high scale, a choice that would, moreover, aggravate the Higgs hierarchy problem~\cite{Vissani:1997ys,Casas:2004gh}. Low-scale realizations can instead exploit an approximate lepton-number symmetry, under which the light-neutrino masses are controlled by a small lepton-number-violating mass parameter $\mu$, such that $m_\nu\simeq\mu\sin^2\theta_N$.\footnote{Since $\mu$ vanishes in the limit of exact lepton-number conservation, its smallness is technically natural, and the observed neutrino masses can arise from small symmetry-breaking effects.} Consequently, sterile neutrinos with masses in the MeV--GeV range can be compatible with the observed neutrino masses while retaining appreciable active--sterile mixing~\cite{Drewes:2016jae,Domcke:2020ety}.

Direct bounds on active--sterile mixing involving the third generation remain comparatively weak, owing to the experimental difficulty of producing and identifying $\nu_\uptau$ states. The best laboratory constraints are from the CHARM beam-dump experiment~\cite{CHARM:1987chp} for sterile-neutrino masses between $10$ and $290$ MeV, based on a search for decays of long-lived sterile neutrinos produced in proton--target collisions~\cite{Orloff:2002de,Boiarska:2021yho}. The resulting limits reach $\sin^2{\theta_N}\sim10^{-4}$ for masses above approximately $160$ MeV, while becoming way weaker at lower masses, above $10^{-1}$ at 10~MeV.  

Complementary bounds arise from cosmology, although they are highly model dependent. In standard thermal histories, active--sterile oscillations can efficiently populate the sterile sector, leading to stringent constraints from both Big Bang nucleosynthesis (BBN) and the cosmic microwave background (CMB). For MeV-scale sterile neutrinos, their production and subsequent decay can modify the expansion rate and weak-interaction rates during BBN, thereby altering the primordial ${}^4$He and deuterium abundances~\cite{Ruchayskiy:2012ksh,Dolgov:2002wy}. Their contribution to the radiation density and the evolution of the neutrino sector is also constrained by measurements of the effective number of relativistic species, $N_{\rm eff}$, from the CMB~\cite{Hansen:2001hi,Vincent:2014rja}. These considerations can therefore exclude sizeable active--sterile mixing for sterile-neutrino masses in the MeV-to-100\,MeV range under consideration.

The situation can change substantially in thermal histories with low reheating temperatures $T_{\rm RH}$. For $T_{\rm RH}$ of a few MeV, sterile-neutrino production through oscillations is suppressed, preventing the sterile sector from reaching thermal equilibrium and significantly relaxing the BBN and CMB constraints~\cite{Gelmini:2008fq,Hasegawa:2020til}. In particular, for $T_{\rm RH}\gtrsim 5\,$MeV and $m_N>\mathcal{O}(1)$ MeV, the sterile-neutrino abundance can be strongly suppressed relative to the standard thermal history, allowing substantially larger active--sterile mixing~\cite{Gelmini:2008fq}.

Additional constraints arise from core-collapse supernovae, in which $N_1$ states can be copiously produced through their mixing with $\nu_\uptau$. Requiring that the energy carried away by sterile neutrinos does not shorten the neutrino burst observed from SN\,1987A excludes $10^{-9}\lesssim\sin^2\theta_N\lesssim10^{-2}$ for $m_N\simeq10\,$MeV~\cite{Mastrototaro:2019vug,Carenza:2023old}, while decays of sterile neutrinos escaping the star constrain mixings down to $\sin^2\theta_N\sim10^{-11}$ for $m_N\lesssim100\,$MeV~\cite{Carenza:2023old,Chauhan:2023sci}. These constraints have been derived for $m_N\gtrsim10\,$MeV, and some of them rest on modeling assumptions that are the subject of ongoing discussion~\cite{Bar:2019ifz,Fiorillo:2023frv}. At $m_N\sim10\,$MeV and $\sin^2\theta_N\gtrsim10^{-2}$, sterile neutrinos are trapped in the core, and the overlap between supernova and laboratory constraints is reduced~\cite{Carenza:2023old}.

\section{\label{sec:exposure} Exposure to sterile neutrinos}

\begin{figure*}[t]
\centering
\includegraphics[width=0.49\textwidth]{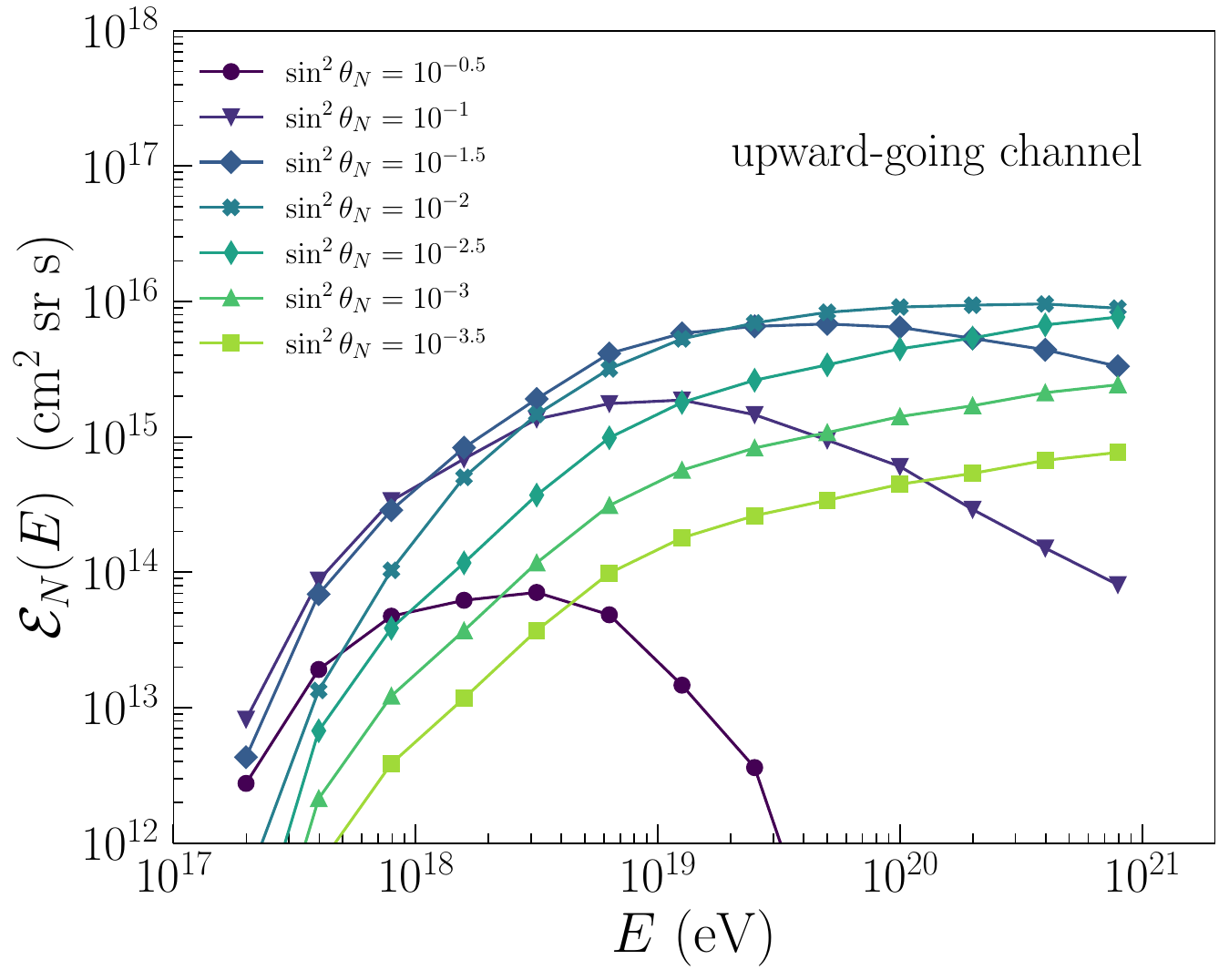}
\includegraphics[width=0.49\textwidth]{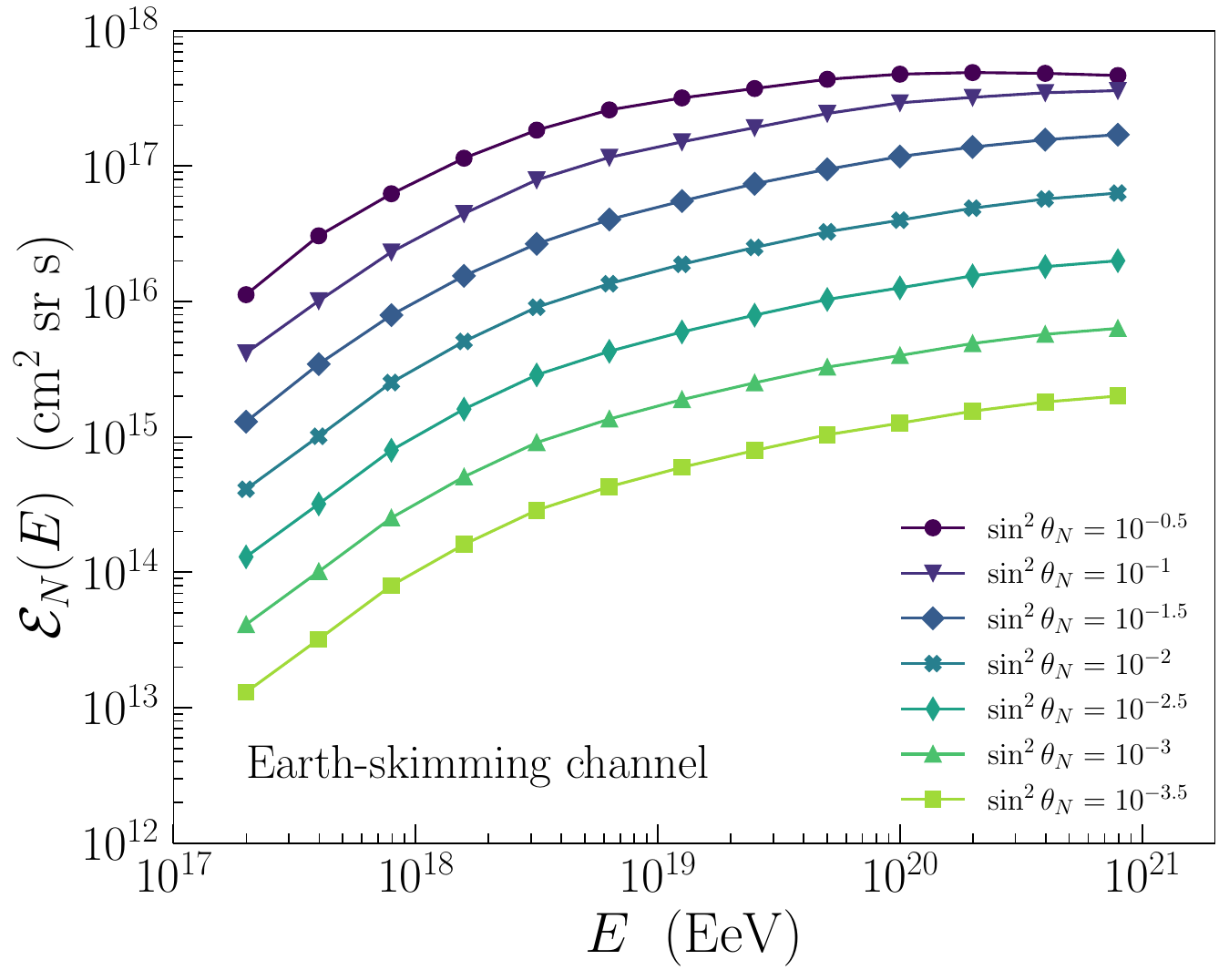}
\caption{Exposure $\mathcal{E}_N(E;\theta_N)$ to air showers induced by MeV-to-100\,MeV-scale sterile neutrinos as a function of energy for several values of the active--sterile mixing angle $\sin^2\theta_N$. Left panel: upward-going channel. Right panel: Earth-skimming channel.}
\label{fig:expo}
\end{figure*}

The exposure to an isotropic flux of sterile neutrinos is obtained by integrating the detection efficiency over time, effective surface area and incident angles $\Omega=(\theta,\varphi)$,  
\begin{equation}
    \label{eqn:exposureN}
    \mathcal{E}_N(E)=\iiint\dif t\,\dif\mathbf{x}\,\dif\Omega\,\left|\cos\theta\right|\,\epsilon_N(E,\mathbf{x},t,\Omega).
\end{equation}
The detection efficiency $\epsilon_N$ to sterile neutrinos in each channel results from that to $\uptau$ leptons that initiate an air shower within the active detection volume, $\epsilon_\uptau$, weighted by the probability that the $\uptau$ decays at an altitude $s\left|\cos\theta\right|$ given a decay length $\lambda(E_\uptau)\simeq 49\,(E_{\uptau}/\mathrm{EeV})\,$km, and convolved with the double-differential number of $\uptau$ leptons induced by incident sterile neutrinos with energy $E$ and angles $\Omega=(\theta,\varphi)$ that emerge at the surface of the Earth with energy $E_\uptau$ and angles $\Omega_\uptau=(\theta_\uptau,\varphi_\uptau)$,
\begin{multline}
    \label{eqn:effN}
    \epsilon_N(E,\mathbf{x},t,\Omega)=\iint\dif\Omega_\uptau\dif E_\uptau\,\frac{\dif^2n_\uptau(E_\uptau,\Omega_\uptau;E,\Omega)}{\dif\Omega_\uptau\dif E_\uptau}\\
    \times \int\dif s\,\frac{\mathrm{e}^{-s/\lambda(E_\uptau)}}{\lambda(E_\uptau)}\epsilon_\uptau(E_\uptau,\Omega_\uptau,s\left|\cos\theta\right|,\mathbf{x},t).
\end{multline}
A full description of the end-to-end methodology used to compute the detection efficiency $\epsilon_\uptau$ is given in Ref.~\cite{PierreAuger:2025hvl} for the upward-going channel, and in Refs.~\cite{PierreAuger:2007vvh,PierreAuger:2019ens} for the Earth-skimming channel. In this study, we use this efficiency extended up to $1000$ EeV. The double-differential number is estimated from Monte Carlo simulations. We generate $n_N$ incident particles crossing the Earth, with their directions sampled uniformly in $\sin^2\theta$ and $\varphi$ according to the integration measure $\dif\Omega\,\left|\cos\theta\right|$ appearing in Eq.~\eqref{eqn:exposureN}. In the upward-going channel, the zenith angle is restricted to $\theta\in[110^\circ,180^\circ]$, while in the Earth-skimming channel we consider $\theta\in[90^\circ,95^\circ]$; in both cases, $\varphi\in[0,360^\circ]$. The subset of realizations in which a $\uptau$ lepton emerges from the Earth is then used to estimate the corresponding double-differential number,
\begin{equation}
\label{eqn:dNdtau}
\frac{\dif^2 n_\uptau}{\dif\Omega_\uptau\dif E_\uptau}=\frac{1}{n_N}\sum_{i=1}^{n_\uptau(E,\Omega;\theta_N)}\delta(E_\uptau-E_{\uptau i})\delta(\Omega_\uptau-\Omega_{\uptau i}),
\end{equation}
where $\delta(x-y)$ denotes the Dirac delta function, and $n_\uptau(E,\Omega;\theta_N)$ is the number of successful realizations for a given incident sterile-neutrino energy and direction. In the Monte Carlo simulation, the density of the Earth is modeled following the \textit{Preliminary
Reference Earth Model}~\cite{Dziewonski:1981xy}. A step-by-step method is used: at each step, the probabilities of the different interactions and of the decay for the $\uptau$ are evaluated as functions of the energy. Both charged-current and neutral-current interactions are taken into account using cross sections reported in Ref.~\cite{Connolly:2011vc} that can be parameterized in the energy range of interest between $10^{-1}$ and $10^{3}~$EeV as
\begin{equation}
    \label{eqn:xs}
    \sigma_{\mathrm{CC}}(E)=2.4\,\sigma_{\mathrm{NC}}(E)=1.05\times 10^{-32}\left(\frac{E}{1~\mathrm{EeV}}\right)^{0.31}\,\mathrm{cm}^{2}
\end{equation}
for active neutrinos, and identical expressions reduced by $\sin^2\theta_N$ for the predominantly-sterile states.  The energy of the outgoing lepton is computed using the inelasticity of the reaction~\cite{Gandhi:1995tf,Connolly:2011vc}. Energy losses of the $\uptau$ can be accounted for under the assumption of continuous energy losses~\cite{Bigas:2008ff}. They include Bremsstrahlung, pair production and deep inelastic scattering. The energy loss model is of the form 
\begin{equation}
    \label{eqn:taulosses}
    -\frac{\dif E_\uptau}{\dif x}\simeq b(E_\uptau)E_\uptau,
\end{equation}
with~\cite{Dutta:2000hh,Armesto:2007tg}
\begin{equation}
b(E_\uptau)=\left(1.2+0.16\,\ln \left(\frac{E_\uptau}{10\,\mathrm{EeV}}\right)\right)\times 10^{-6}\,\mathrm{g}^{-1}\,\mathrm{cm}^2.
\end{equation}
The TAUOLA package~\cite{Jadach:1993hs} is used to simulate $\uptau$ decay and obtain the secondary particles and their energies. On inserting Eq.~\eqref{eqn:dNdtau} into Eq.~\eqref{eqn:effN}, and on carrying out the integrations, the detection efficiency reads as
\begin{multline}
    \label{eqn:effNbis}
    \epsilon_N(E,\mathbf{x},t,\Omega;\theta_N)=\\
    \frac{1}{n_N}\sum_{i=1}^{n_\uptau}\int\dif s\,\frac{\mathrm{e}^{-\frac{s}{\lambda(E_{\uptau i})}}}{\lambda(E_{\uptau i})}\epsilon_\uptau(E_{\uptau i},\Omega_{\uptau i},s\left|\cos\theta_{\uptau i}\right|,\mathbf{x},t),
\end{multline}
where the dependence on $E$, $\Omega$, and $\theta_N$ is encoded in the number of successful Monte Carlo realizations, $n_\uptau\equiv n_\uptau(E,\Omega;\theta_N)$, which controls the sum over $i$ (cf. Eq.~\eqref{eqn:dNdtau}). Finally, to plug Eq.~\eqref{eqn:effNbis} into Eq.~\eqref{eqn:exposureN}, we use the Monte Carlo estimation of the position- and time-averaged detection efficiency, denoted below as $\langle \epsilon_\uptau\rangle_{\mathbf{x},t}$, integrated over total duration $\Delta T=14\,$yr ($\Delta T=17\,$yr) and area $S=10{,}000\,$km$^2$  ($S=3{,}000\,$km$^2$) in the case of the upward-going channel~\cite{PierreAuger:2025hvl} (Earth-skimming channel~\cite{PierreAuger:2019ens}),
\begin{multline}
    \label{eqn:exposureNbis}
    \mathcal{E}_N(E;\theta_N)=\frac{S\,\Delta T}{n_N}\int\dif\Omega\,\left|\cos\theta\right|\\
    \times\sum_{i}^{n_\uptau} \int\dif s\,\frac{\mathrm{e}^{-s/\lambda(E_{\uptau i})}}{\lambda(E_{\uptau i})}\langle\epsilon_\uptau(E_{\uptau i},\Omega_{\uptau i},s\left|\cos\theta_{\uptau i}\right|)\rangle_{\mathbf{x},t}.
\end{multline}
The last integration over solid angle yields a global factor $\pi(1-\cos^2 110^\circ)$ in the upward-going channel ($\pi(\cos^2 95^\circ)$ in the Earth-skimming one), inherited from the scheme to draw particles at random when conducting the Monte Carlo simulation to estimate $\dif^2n_\uptau/\dif\Omega_\uptau\dif E_\uptau$. 

The exposure is shown in Fig.~\ref{fig:expo} as a function of energy for several values of $\sin^2\theta_N\in[10^{-3},10^{-0.5}]$. In the upward-going channel (left panel), the exposure increases rapidly as $\sin^2\theta_N$ decreases from $10^{-0.5}$ to $10^{-2}$, reflecting the transition from an opaque to a transparent Earth. For sufficiently small mixing, the sterile neutrino traverses the Earth essentially without attenuation, and the exposure becomes proportional to the interaction probability, yielding the asymptotic scaling $\mathcal{E}_N\propto\sin^2\theta_N$. The maximum sensitivity is reached in the intermediate regime, $\sin^2\theta_N\sim10^{-1}$--$10^{-2.5}$, where the interaction length is comparable to the characteristic chord length through the Earth. In the Earth-skimming channel (right panel), the behavior is both quantitatively and qualitatively different. Because the Earth-skimming geometry already corresponds to near-optimal propagation conditions for Standard-Model cross sections, the exposure remains close to its maximal value down to $\sin^2\theta_N\sim10^{-1}$. For smaller mixings, sterile neutrinos increasingly enter the transparent regime and become less likely to convert into observable $\uptau$ leptons within the narrow layer of crust accessible to this channel. Consequently, the exposure decreases approximately linearly with $\sin^2\theta_N$, but remains systematically larger than that of the upward-going channel by roughly one order of magnitude.

Several sources of systematic effects affect the exposure calculation. Actually, they inherit from those pertaining to Standard Model $\uptau$ neutrinos~\cite{PierreAuger:2015ihf}. The dominant systematic uncertainties, $^{+34\%}_{-28\%}$, arise from the modeling of the $\uptau$-lepton energy losses during propagation through the Earth that directly determine the energy of the emerging $\uptau$ and therefore the detector exposure. Additional contributions are negligible in comparison; they originate from the simulation of extensive air showers, primarily through the choice of shower-propagation code and hadronic interaction model, and from the topography surrounding the Observatory.

\section{\label{sec:fluxlimits} Flux upper limits}

The search for upward-going air showers with zenith angles larger than $110^\circ$ and energies above $0.1$~EeV using the Fluorescence Detector of the Observatory yielded one candidate event, consistent with an expected background of $0.27 \pm 0.12$ misreconstructed cosmic-ray showers over the search period from January 1, 2004, to December 31, 2018~\cite{PierreAuger:2025hvl}. On the other hand, the latest search for Earth-skimming events over the period from January 1, 2004, to December 31, 2021 yielded no candidates~\cite{PierreAuger:2023pjg}. These observations can be translated into upper limits on the flux of sterile neutrinos at $90\%$ confidence level. To this end, we compute the expected number of signal events as
\begin{equation}
    n_{\mathrm{exp}}(\theta_N)=\int \mathrm{d}E \, \Phi_N(E)\,\mathcal{E}_N(E;\theta_N),
\end{equation}
where $\Phi_N(E)$ is the differential flux of $N_1$ particles. The upper limit is then obtained by requiring $n_{\mathrm{exp}}(\theta_N)=4.09$ $(n_{\mathrm{exp}}(\theta_N)=2.44)$, appropriate for one (zero) observed event in the presence of the estimated background~\cite{Feldman:1997qc}. Given the absence of signal events, the limits are largely dominated by the exposure of the Earth-skimming channel. 

\begin{figure}[t]
\centering
\includegraphics[width=0.5\textwidth]{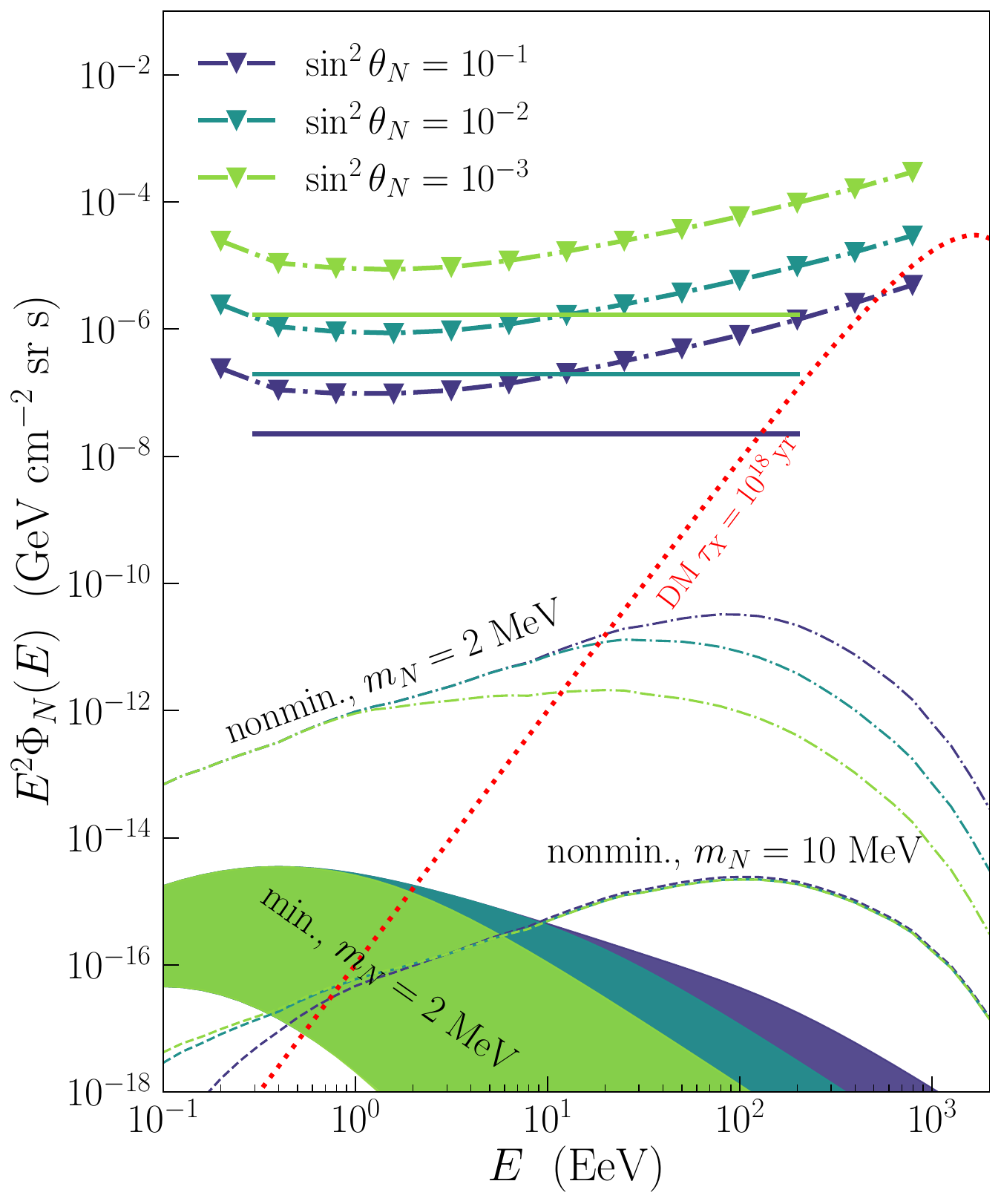}
\caption{Upper limits at 90\% confidence level on the diffuse flux of MeV-to-100\,MeV-scale sterile neutrinos. Results are shown for three benchmark values of the active--sterile mixing angle, $\sin^2\theta_N=10^{-1}$, $10^{-2}$, and $10^{-3}$. Triangle symbols connected by dot-dashed lines indicate differential limits, while solid lines show the corresponding integral limits. For comparison, the colored bands and dashed-lines show benchmark cosmogenic sterile-neutrino fluxes obtained from active--sterile oscillations of astrophysical neutrinos during propagation over cosmological distances. The red dotted curve illustrates the sterile-neutrino flux expected from the decay of a superheavy dark-matter particle into sterile neutrinos with mass $M_\chi=10^{13}\,$GeV and  lifetime $\tau_\chi=10^{18}\,$yr.}
\label{fig:fluxlimits}
\end{figure}

For definiteness, we assume an isotropic flux with a benchmark spectral shape proportional to $E^{-2}$, as commonly adopted in searches for high-energy neutrinos. The resulting upper limits on $\Phi_N$ are shown in Fig.~\ref{fig:fluxlimits}. Results for $\sin^2\theta_N=10^{-1}$, $10^{-2}$, and $10^{-3}$ are displayed as dark blue, cyan, and green triangles, respectively, with interpolating lines in the conventional $E^2\Phi_N$ representation. Both differential (triangle symbols joined by dashed lines) and integral (plain lines) limits are shown. For completeness, the limits are also reported in Table~\ref{tab:limits} for $\sin^2\theta_N=10^{-1}$. Overall, for mixing angles $\sin^2\theta_N \lesssim 10^{-1}$, we obtain with 90\% confidence level the bound over $[0.3,200]~$EeV
\begin{equation}
    \label{eq:bound}
    E^2\Phi_N\leq 2\times 10^{-9}\sin^{-2}\theta_N\,\,\mathrm{GeV}\,\mathrm{cm}^{-2}\,\mathrm{sr}^{-1}\,\mathrm{s}^{-1}.
\end{equation}

\begin{table}[t]
\caption{90\% confidence level upper limits on $E^2\Phi_N(E)$, applicable for $\sin^2\theta_N<10^{-1}$.}
\begin{ruledtabular}
\begin{tabular}{l c}
$\log_{10}{(E/\mathrm{eV})}$ & $E^2\Phi_N^{90\%}$ (GeV\,cm$^{-2}$\,sr$^{-1}$\,s$^{-1}$)\\
\colrule
&\\[-0.8em]
17.15 & $2.4\times10^{-7}\sin^{-2}\theta_N$ \\
17.45 & $1.1\times10^{-7}\sin^{-2}\theta_N$ \\
17.75 & $9.8\times10^{-8}\sin^{-2}\theta_N$ \\
18.05 & $9.7\times10^{-8}\sin^{-2}\theta_N$ \\
18.35 & $1.1\times10^{-7}\sin^{-2}\theta_N$ \\
18.65 & $1.4\times10^{-7}\sin^{-2}\theta_N$ \\
18.95 & $2.0\times10^{-7}\sin^{-2}\theta_N$ \\
19.25 & $3.1\times10^{-7}\sin^{-2}\theta_N$ \\
19.55 & $4.9\times10^{-7}\sin^{-2}\theta_N$ \\
19.85 & $8.1\times10^{-7}\sin^{-2}\theta_N$ \\
20.15 & $1.4\times10^{-6}\sin^{-2}\theta_N$ \\
20.45 & $2.6\times10^{-6}\sin^{-2}\theta_N$ \\
20.75 & $4.9\times10^{-6}\sin^{-2}\theta_N$ 
\end{tabular}
\end{ruledtabular}

\label{tab:limits}
\end{table}

For comparison, we also show, using the same color coding for $\sin^2\theta_N$, benchmark cosmogenic expectations for $\Phi_N$ arising from oscillations of the $\nu_\uptau$ flux. In the standard scenario, only electron and muon neutrinos are produced in cosmic-ray interactions at the sources or during propagation to Earth. However, neutrino oscillations convert a substantial fraction of this flux into $\nu_\uptau$ by the time it reaches Earth. In the $3+1$ framework considered here, a small sterile component is likewise generated through flavor transitions,
\begin{equation}
    \label{eq:PhiN_general}
    \Phi^0_N(E)=\sum_{\alpha=\elec,\upmu,\uptau,\st} P_{\alpha\to \st}\,\Phi_\alpha^0(E),
\end{equation}
where $\Phi_\alpha^0(E)$ denotes the source flux in flavor state $\alpha$. For propagation over cosmological distances, the oscillation lengths are negligible compared to the baseline, and the oscillatory terms average to zero. The flavor-transition probabilities therefore reduce to their phase-averaged values,
\begin{equation}
    \label{eq:Pavg}
    P_{\alpha\to \st}=\sum_{i=1}^{4}|U_{\alpha i}|^2 |U_{\st i}|^2.
\end{equation}
In practice, we begin with the standard $3\times3$ PMNS sector, adopting the current best-fit three-flavor mixing parameters~\cite{Esteban:2024eli}. We then introduce a sterile--$\uptau$ rotation in flavor space with mixing angle $\theta_N$. To leading order in $\sin^2\theta_N$, the sterile-neutrino flux is
\begin{equation}
    \label{eq:PhiN_leading}
    \Phi^0_N(E)=\sin^2\theta_N
    \sum_{\alpha=\elec,\upmu,\uptau}|U_{\alpha3}|^2\,\Phi_\alpha^0(E).
\end{equation}
Assuming the canonical pion-decay source composition, $\Phi_\elec^0:\Phi_\upmu^0:\Phi_\uptau^0 = 1:2:0$, Eq.~\eqref{eq:PhiN_general} reduces to
\begin{equation}
    \label{eq:PhiN_pion}
    \Phi^0_N(E)=\sin^2\theta_N
    \left(
    \frac{1}{3}|U_{\elec 3}|^2+\frac{2}{3}|U_{\upmu3}|^2
    \right)\Phi^0(E),
\end{equation}
where $\Phi^0(E)$ is the total neutrino flux at production. 

Propagation of the sterile state over cosmological distances is affected by its decay. For a sterile neutrino of mass $m_N$ below 100~MeV, the dominant decay channel is $N_1\to\nu_\uptau\nu_i\overline{\nu}_i$, the width of which reads as~\cite{Dicus:1977nn,Dicus:1977qy,Roncadelli:1981tn,Hosotani:1981mq,Barger:1995ty,Gorbunov:2007ak}
\begin{equation}
\Gamma_N\simeq 1.5~10^{2}
\left(\frac{\theta_N^2}{10^{-4}}\right)
\left(\frac{m_N}{100\,\mathrm{MeV}}\right)^5\,\mathrm{s}^{-1}.
\end{equation}
In the laboratory frame, the decay length consequently reads as
\begin{equation}
\lambda_N(E)\simeq
2\left(\frac{10^{-2}}{\theta_N^2}\right)
\left(\frac{10\,\mathrm{MeV}}{m_N}\right)^6
\left(\frac{E}{1\,\mathrm{EeV}}\right)~\mathrm{kpc}.
\label{eq:decaylength}
\end{equation}
The mass range $m_N=\mathcal{O}(\mathrm{MeV})$ provides a compromise between propagation considerations and cosmological constraints, especially in the context of low-reheating temperatures to relax the otherwise stringent BBN and CMB constraints on active--sterile mixing. In particular, the range $10^{-2}\lesssim\sin^2\theta_N\lesssim 10^{-1}$, which is the optimal one in terms of detection prospects, remains viable for $m_N\gtrsim 2\,$MeV~\cite{Gelmini:2008fq,Hasegawa:2020til}. At the same time, the strong mass dependence of the decay length in Eq.~\eqref{eq:decaylength} ensures that sterile-neutrino fluxes are not completely attenuated over cosmological propagation distances throughout the parameter space of interest. Masses substantially above this range lead to increasingly rapid decays and hence to strong attenuation of the UHE sterile-neutrino flux, whereas masses below the MeV scale are more strongly constrained in the standard cosmological history and are less relevant in the low-reheating-temperature scenarios.

The decay depletes the sterile-neutrino flux at a given energy while producing $\uptau$ neutrinos at lower energies. During their subsequent propagation, these daughter neutrinos can oscillate back into the sterile state, thereby regenerating the sterile-neutrino flux. However, the regenerated sterile neutrinos rapidly accumulate at energies below the range of interest. The effect of regeneration can therefore be neglected for the flux in the energy range considered here, and the propagated flux from a source at redshift $z$ can be approximated by an attenuated one with distance, accounting for the expansion of the Universe,
\begin{multline}
\Phi_N(E,z)
\simeq
\Phi_N^0\!\left(E(1+z)\right)\times\\
\exp\left(
-\frac{m_N\Gamma_N}{E}
\int_0^z
\frac{dz'}{(1+z')^2 H(z')}
\right),
\label{eq:PhiN_decay}
\end{multline}
where $H$ denotes the Hubble rate, which, in the redshift range of interest, reads as $H(z)=H_0\sqrt{\Omega_\mathrm{m}(1+z)^3+\Omega_\Lambda}$, with $H_0\simeq 73.5$\,km\,s$^{-1}$\,Mpc$^{-1}$ the Hubble constant, $\Omega_\mathrm{m}\simeq 0.3$ and $\Omega_\Lambda\simeq0.7$~\cite{H0DN2026}.

To estimate $\Phi^0(E)$, we adopt cosmogenic flux models consistent with our measurements of the energy spectrum and mass composition, see, e.g., Refs.~\cite{PierreAuger:2022atd,Aloisio:2013hya,Biehl:2017hnb,AlvesBatista:2018zui,Boncioli:2018lrv,Zhang:2018agl,Condorelli:2022vfa,Muzio:2022bak,Rodrigues:2020pli,Muzio:2023skc,Berat:2024rvf}. The filled bands correspond to sets of predictions compiled in Ref.~\cite{Berat:2024rvf} obtained within minimal source scenarios, in which a single extragalactic population dominates above the ankle energy, and corrected for effective attenuation effects for $m_N=2\,$MeV. A conservative estimate stems from the production \textit{en route} alone. For various source evolutions, there is convergence in the literature for a (single-flavor) flux scaling as $E^2 \Phi_\nu \simeq 2 \times 10^{-10}\,$GeV\,cm$^{-2}$\,sr$^{-1}$\,s$^{-1}$ at 100\,PeV and decreasing in the energy range of interest for this study down to $E^2 \Phi_\nu \simeq 10^{-12}\,$GeV\,cm$^{-2}$\,sr$^{-1}$\,s$^{-1}$ at $1\,$EeV; specifically, we follow the GRB-like source evolution of Ref.~\cite{AlvesBatista:2018zui}. When considering specifics of in-source production such as e.g. tidal disruption events~\cite{Boncioli:2018lrv}, low-luminosity gamma-ray bursts~\cite{Zhang:2018agl} or starburst-galaxy environments~\cite{Condorelli:2022vfa}, the expected fluxes reach $E^2 \Phi_\nu \simeq 2\times 10^{-9}\,$GeV\,cm$^{-2}$\,sr$^{-1}$\,s$^{-1}$ at the peak energy 100\,PeV; we follow here Ref.~\cite{Muzio:2022bak}. We show also enhanced flux scenarios including an additional ultrahigh-energy proton component~\cite{Muzio:2023skc} as the dotted-dashed curves (effective attenuation effects for $m_N=2\,$MeV) and the dashed curves (effective attenuation effects for $m_N=10\,$MeV). Attenuation effects remain small for $m_N=2\,$MeV at high energies, while they gradually become relevant below $\sim 100\sin^2\theta_N\,$EeV. For $m_N=10\,$MeV, in contrast, the much shorter decay length leads to substantial attenuation over the relevant propagation distances. Coincidentally, the resulting propagated fluxes become nearly independent of $\sin^2\theta_N$ in the energy range of interest. This behavior results from a compensation between two competing effects: decreasing $\sin^2\theta_N$ reduces the sterile-neutrino production rate, and hence the initial flux, but simultaneously increases the decay length, thereby reducing the attenuation during propagation. 

All of these benchmark predictions lie several orders of magnitude below the current experimental sensitivity, underscoring the difficulty of probing cosmogenic sterile-neutrino fluxes with present instruments.

Finally, we also display a representative flux prediction arising from the decay of a specific dark-matter candidate, discussed in the next section.

\section{\label{sec:shdm} Constraints on active--sterile mixing angle from superheavy dark matter decay}

In Fig.~\ref{fig:fluxlimits}, a flux from a dark-matter model is displayed as a dotted line, consistent with current constraints but potentially accessible with increased future exposure, assuming a hypothetical superheavy particle $\chi$ decaying into sterile neutrinos alone~\cite{Merle:2013wta,Heurtier:2019git,Morrison:2022zwk}. A dark-matter candidate that couples exclusively to sterile neutrinos would be \textit{a priori} particularly challenging to probe with conventional searches. For such a scenario sometimes referred to as a ``$\nu$ nightmare''~\cite{Morrison:2022zwk}, the detection channels described in this work provide unique sensitivity to probe the  sterile-neutrino flux at UHE. To realize this scenario in a generic way, we consider here a scalar particle $\chi$ and an effective interaction term that leads to the production of $k$ pairs of sterile neutrinos,
\begin{eqnarray}
    \label{eqn:Leff}
    \mathcal{L}_\mathrm{eff}=\frac{g}{\Lambda^{3k-3}}\chi \overline{({N_1)^\mathrm{c}}}N_1\dots \overline{({N_k)^\mathrm{c}}}N_k+\mathrm{h.c.},
\end{eqnarray}
with $g$ a dimensionless constant and $\Lambda$ a high-energy scale. Here $N_1$ denotes as previously the  predominantly-sterile state relevant for propagation through the Earth, while $N_{2,\dots,k}$ are additional sterile Majorana states, the detailed flavor composition of which is irrelevant for the phenomenology considered below. An ultraviolet completion of this interaction that would protect the $\chi$ particle from additional decay channels is beyond the scope of the present study. Instead, we follow the customary approach of constraining a specific decay channel by using Eq.~(\ref{eqn:Leff}) as a phenomenological parametrization of superheavy dark matter decay into sterile neutrinos, with $k$ treated as a free parameter. While the minimal operator corresponds to $k=1$, larger values of $k$ naturally suppress the decay width through additional powers of the heavy scale $\Lambda$, potentially allowing for $\tau_\chi$ much longer than the age of the Universe even for dimensionless couplings of order unity.

Considering the Galactic contribution alone, the flux $\Phi_N(E)$ is governed by the energy density of dark matter $\rho_\chi$ integrated along the line of sight, the energy spectrum of the decay byproducts, the mass $M_\chi$ of the dark matter candidate and its lifetime $\tau_\chi$, 
\begin{multline}
    \label{eqn:phiDM}
    \Phi_{N_1}(E,\Omega)=\frac{1}{4\pi M_\chi\tau_\chi}\frac{\dif n_{N_1}(E)}{\dif E}\\
    \times\int_0^\infty\dif s\,\rho_\chi(\mathbf{x}_\odot+s\mathbf{n})\exp{\left(-\frac{s}{\lambda_{N_1}(E)}\right)},
\end{multline}
with $\mathbf{n}\equiv \mathbf{n}(\Omega)$ the unit vector on the sphere. The energy spectrum associated to each $N_1$ particle is $\dif n_{N_1}/\dif E=\delta(E-M_\chi/2)$ for $k=1$, while it reads as (see Appendix~\ref{app:sterile_spectrum}) 
\begin{eqnarray}
    \label{eqn:dnNdE}
    \frac{\dif n_{N_1}}{\dif E}=\frac{4}{B(k,3k-3)M_\chi}\left(\frac{2E}{M_\chi}\right)^2\left(1-\frac{2E}{M_\chi}\right)^{3k-4},
\end{eqnarray}
for $k\geq 2$, with $B(x,y)$ the Euler beta function. For illustrative purposes, the flux expected for $k=4$ and $\tau_\chi=10^{18}\,\mathrm{yr}$ is shown as the dotted line after averaging over arrival directions. For definiteness, the mass of the decaying particle is fixed at $M_\chi = 10^{13}\,\mathrm{GeV}$. Such a superheavy dark matter candidate can be produced through minimal gravitational interactions during the reheating epoch, yielding relic abundances consistent with the observed dark matter density without requiring sizable couplings to the Standard Model (see, e.g., Ref.~\cite{Chung:2001cb}). In these scenarios, quantum fluctuations of the scalar field generated during inflation induce primordial isocurvature perturbations in the dark matter density. Measurements of the CMB place stringent limits on such isocurvature fluctuations, which in turn constrain the viable mass range for gravitationally produced scalar dark matter candidates. Nevertheless, for reasonable inflationary scales, masses around $M_\chi \sim 10^{13}\,\mathrm{GeV}$ remain compatible with current cosmological constraints~\cite{Chung:2004nh,Garcia:2023awt}. 

\begin{figure}[!t]
\centering
\includegraphics[width=0.5\textwidth]{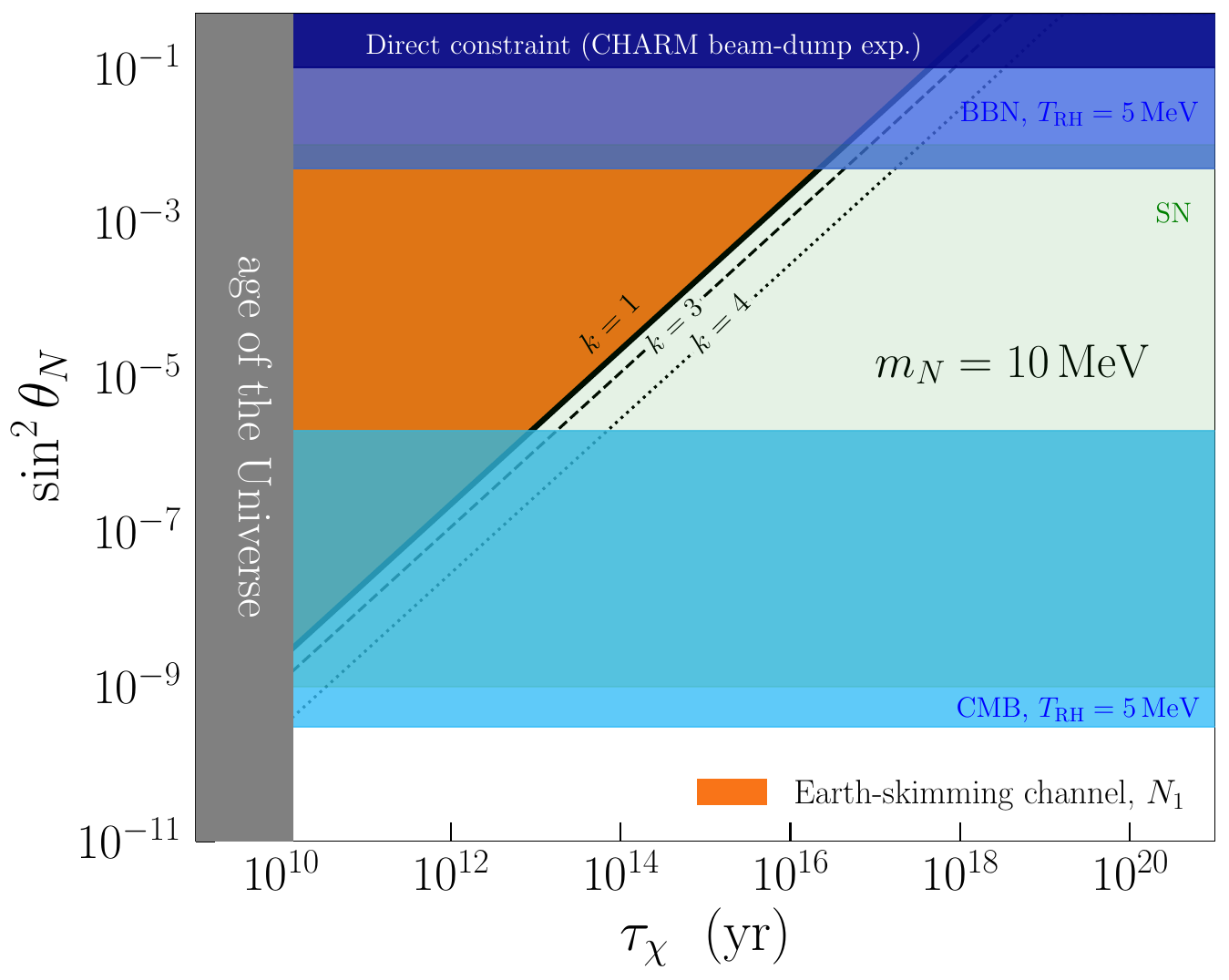}
\caption{Constraints at 90\% confidence level in the $(\sin^2\theta_N,\tau_\chi)$ plane for a superheavy dark-matter particle of mass $M_\chi=10^{13}\,$GeV decaying exclusively into sterile neutrinos. The orange region is excluded by the air-shower searches presented in this study assuming a decay multiplicity $k=1$. For comparison, the dotted and dashed curves show the corresponding envelopes for $k=3$ and $k=4$, respectively. Also shown are existing constraints on the active--sterile mixing angle from the CHARM experiment, from cosmological observations in a scenario with a reheating temperature $T_\mathrm{RH}=5\,$MeV, and from supernova.}
\label{fig:theta2vstau-10MeV}
\end{figure}
 
Using the benchmark values $M_\chi=10^{13}\,$GeV, the number of expected events depends on both $\sin^2\theta_N$ and $\tau_\chi$. Therefore, pairs $(\sin^2\theta_N, \tau_\chi)$ can be excluded at 90\% confidence level by requiring  the expected number of events in the Earth-skimming channel $n_\mathrm{exp}(\theta_N,\tau_\chi)$ to exceed 2.44. Although the exposure is available from simulations up to 1000\,EeV, we consider sterile-neutrino energies up to $M_\chi/2=5000$\,EeV. To do so, we extrapolate the exposure using its asymptotic behavior. We have verified that this extrapolation has a negligible impact on the results because, for $k\geq2$, the decay spectrum is strongly suppressed near the endpoint and contributes only marginally above 1000\,EeV. 

We first consider the constraints for a sterile-neutrino mass $m_N=10\,$MeV, for which the decay length is sufficiently large to allow propagation over Galactic distances at the energies of interest. The resulting exclusion region is shown in orange in Fig.~\ref{fig:theta2vstau-10MeV}. For $\sin^2\theta_N$ around $10^{-2}$, the lifetime of the superheavy dark-matter particle is constrained to be $\tau_\chi\gtrsim10^{16}\,\mathrm{yr}$ for $k=1$. At smaller mixing angles, the lower bound on $\tau_\chi$ weakens  linearly with $\sin^2\theta_N$, reflecting the reduced interaction probability of the sterile neutrinos as the Earth becomes increasingly transparent to them. For illustration, we also show the corresponding limits for $k=3$ and $k=4$ as dotted and dashed lines, respectively The constraints become stronger for larger values of $k$, corresponding to a greater multiplicity of the decay products and hence to a larger sterile-neutrino flux.  Overall, the constraints obtained on $\tau_\chi$ are weaker than those typically derived for superheavy particles, namely $\tau_\chi\gtrsim 10^{23}\,$yr~\cite[see, e.g.,][]{Ishiwata:2019aet,Alcantara:2019sco,Berat:2022iea,IceCube:2022clp,PierreAuger:2023vql,Deligny:2024fyx}. This can be traced to several key features of the scenario under consideration, in particular to the decay products that consist dominantly of sterile neutrinos\footnote{A subdominant contribution arises from $\nu_\uptau$-induced air showers generated by sterile-neutrino oscillations, with $\Phi_{\nu_\uptau}\simeq \sin^2\theta_N(1+|U_{\uptau3}|^2)\Phi_N$. However, this contribution has a negligible impact on the exclusion limits, since the corresponding exposure is suppressed by an additional $\sin^2\theta_N$ factor.} with low multiplicity and hard energy spectra.  

Independently of the specific dark-matter production scenario, the parameter space is restricted by existing bounds on the active--sterile mixing. As discussed in Section~\ref{sec:theory}, the CHARM beam-dump experiment provides the only direct, laboratory-based constraint (dark-blue region). Cosmological and supernova observations provide complementary constraints, shown as the navy-blue, light-blue, and light-green regions for BBN, CMB, and supernovae, respectively. Although more stringent, these constraints are indirect and rely on additional assumptions. The cosmological ones are evaluated for a low-reheating-temperature scenario with $T_\mathrm{RH}=5\,$MeV~\cite{Gelmini:2008fq}; for higher reheating temperatures, sterile-neutrino production in the early Universe is enhanced and the corresponding constraints become substantially stronger. Supernova constraints exclude a large part of the region considered here, but they rely on the modeling of the stellar core and of the sterile-neutrino emission. Our bounds provide an independent probe, based on the interactions of sterile neutrinos within the Earth and subject to entirely different systematic uncertainties.  

To explore a different regime of the parameter space, we also consider $m_N=100\,$MeV, for which the sterile-neutrino decay length is much shorter than Galactic distances. In this case, the sterile-neutrino flux is efficiently converted into lower-energy $\uptau$ neutrinos through $N_1$ decay. The resulting $\uptau$-neutrino flux can be written as
\begin{multline}
\label{eqn:phiDMbis}
\Phi_{\nu_\uptau}(E_\nu,\Omega)=
\int_0^\infty\dif s\,
\frac{\rho_\chi(\mathbf{x}_\odot+s\mathbf{n})}
{4\pi M_\chi\tau_\chi}
\left[1-\mathrm{e}^{-s/\lambda_{N_1}(E_\nu)}\right]\\
\times
\int\frac{\dif y}{1-y}
\frac{\dif n_{N_1}\left(\frac{E_\nu}{1-y}\right)}
{\dif E_\nu}
\frac{\dif n_{\nu_\uptau}\left(\frac{E_\nu}{1-y},y\right)}
{\dif y},
\end{multline}
where $\dif n_{\nu_\uptau}/\dif y$ denotes the energy distribution of $\uptau$ neutrinos among the decay products of $N_1$ in the channels $N_1\to\nu_\uptau\nu_i\bar\nu_i$ and $N_1\to\bar\nu_\uptau\bar\nu_i\nu_i$. In the limit $m_N\ll m_Z$, the three-body decay $N\to3\nu$ is similar to the ordinary leptonic (Michel) decays, neglecting the masses of the final-state neutrinos. In the rest frame of $N_1$, defining $x=2E_\nu^*/m_N$, the energy distributions can be expressed in terms of two hard and one soft spectra, $f_\mathrm{h}(x)=2x^2(3-2x)$ and $f_\mathrm{s}(x)=12x^2(1-x)$. We neglect polarization effects and use the angle-integrated decay spectra, corresponding to an unpolarized $N_1$ ensemble. For $E_N\gg m_N$, the laboratory-frame energy is related to the rest-frame variables by $E_\nu/E_N\simeq x(1+\cos\theta^*)/2$. 
Introducing $y=E_\nu/E_N$ and integrating over the decay angle gives the corresponding laboratory-frame spectra
\begin{eqnarray}
\frac{\dif n_{\mathrm{h}}}{\dif y}
&=&\frac{5}{3}-3y^2+\frac{4}{3}y^3,
\\
\frac{\dif n_{\mathrm{s}}}{\dif y}
&=&2-6y^2+4y^3.
\end{eqnarray}
The $\uptau$-neutrino spectrum is then
\begin{equation}
\frac{\dif n_{\nu_\uptau}}{\dif y}
=\frac{4}{3}\frac{\dif n_{\mathrm{h}}}{\dif y}
+\frac{1}{3}\frac{\dif n_{\mathrm{s}}}{\dif y},
\end{equation}
where the coefficients account for both the multiplicity of the hard and soft spectra and the $1/3$ flavor projection of the $\nu_i\bar\nu_i$ pair onto the $\uptau$ flavor.

\begin{figure}[t]
\centering
\includegraphics[width=0.5\textwidth]{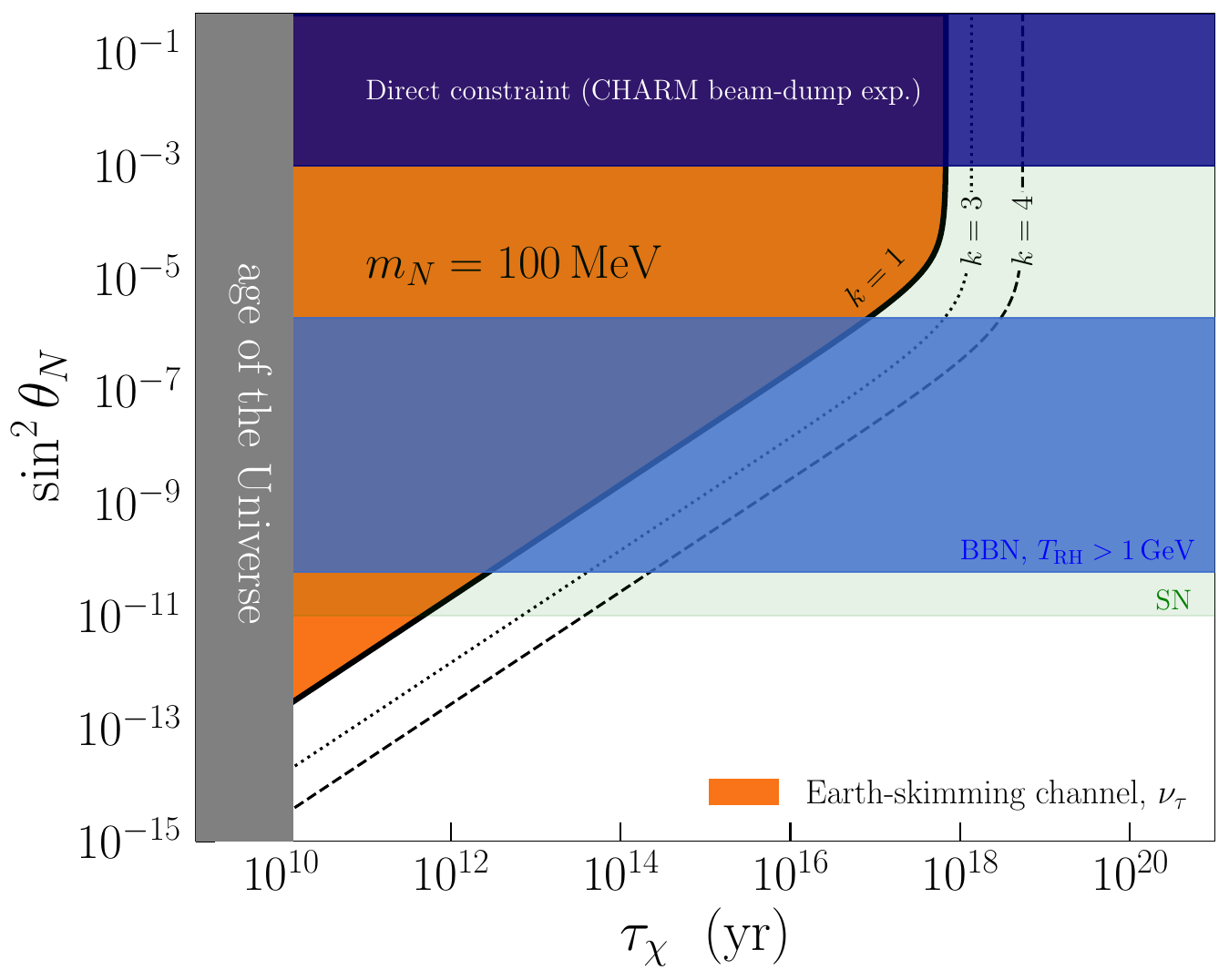}
\caption{ Same as Fig.~\ref{fig:theta2vstau-10MeV}, for $m_N=100\,$MeV.}
\label{fig:theta2vstau-100MeV}
\end{figure}

The resulting exclusion region is shown in orange in Fig.~\ref{fig:theta2vstau-100MeV}. For $m_N=100\,$MeV, the CHARM constraint (dark-blue region) excludes mixing angles around $\sin^2\theta_N\simeq10^{-3}$. Cosmological constraints are evaded in the low-reheating-temperature scenario and become relevant only for higher reheating temperatures (medium-blue region, labeled as ``BBN, $T_\mathrm{rh}>1\,$GeV''), while supernova constraints (light-green region, labeled as ``SN'') extend to   $10^{-11}\lesssim\sin^2\theta_N\lesssim10^{-3}$. Our bounds extend to even smaller values for lifetimes still larger the age of the Universe. 

For heavier sterile neutrinos, supernova constraints fade away above a few hundred MeV, as production in the core becomes Boltzmann-suppressed, and cosmological constraints weaken as the sterile-neutrino lifetime shortens. Our constraints, in contrast, are expected to remain approximately stable; the only remaining bound on the mixing in this mass range is then that from DELPHI, $\sin^2\theta_N\lesssim10^{-5}$~\cite{DELPHI:1996qcc}. Above the pion threshold, however, the opening of the $N_1\to\pi^0\nu_\uptau$ channel, followed by $\pi^0\to\gamma\gamma$, would yield an ultrahigh-energy gamma-ray flux, so that the dark-matter lifetime would then be more strongly constrained by photon searches~\cite{PierreAuger:2023vql}.

\section{\label{sec:discussion} Summary}

We have investigated the sensitivity of the Pierre Auger Observatory to light sterile neutrinos with masses in the MeV-to-100\,MeV range, focusing on a benchmark $3+1$ scenario in which a single predominantly sterile state mixes mainly with the $\nu_\uptau$ flavor. The $N_1$  neutrino interacts with matter through its suppressed active component and can therefore traverse substantial chord lengths through the Earth before producing a $\uptau$ lepton that emerges and decays in the atmosphere. For the mass and energy ranges considered here, matter effects on the active--sterile mixing are negligible, so that the interaction probability is governed by the vacuum mixing parameter $\sin^2\theta_N$.

We have calculated the exposure to sterile neutrinos for both the Earth-skimming and upward-going air-shower channels using a detailed Monte Carlo treatment of propagation through the Earth, neutrino interactions, $\uptau$ energy losses, and $\uptau$ decay. The Earth-skimming channel provides the larger exposure over most of the parameter space, while the upward-going channel is most sensitive for intermediate mixings, where the sterile-neutrino interaction length is comparable to the characteristic chord length through the Earth.

Using the observed event samples, we have derived upper limits at $90\%$ confidence level on an isotropic diffuse sterile-neutrino flux. For an $E^{-2}$ spectrum and $\sin^2\theta_N\lesssim10^{-1}$, the resulting limit can be expressed approximately as
\begin{equation}
E^2\Phi_N\lesssim
2\times10^{-9}\sin^{-2}\theta_N\,
\mathrm{GeV\,cm^{-2}\,sr^{-1}\,s^{-1}},\nonumber
\end{equation}
over the energy interval $0.3$--$200$ EeV. The limits are dominated by the Earth-skimming channel. Benchmark sterile-neutrino fluxes obtained from active--sterile oscillations remain several orders of magnitude below the present  sensitivity.

We have also considered sterile neutrinos produced in the decay of a superheavy dark-matter particle, assuming that the dark matter decays exclusively into sterile states. For $M_\chi=10^{13}$ GeV, the air-shower search constrains the dark-matter lifetime as a function of the active--sterile mixing. For $m_N=10$ MeV, the sterile flux can propagate over Galactic distances, while for $m_N=100$ MeV, sterile-neutrino decay efficiently transfers the injected energy into lower-energy $\nu_\uptau$ neutrinos that can themselves induce air showers. The resulting constraints must be considered together with existing direct and cosmological bounds on active--sterile mixing, the latter depending strongly on the assumed reheating temperature.

Overall, this study demonstrates that ultrahigh-energy air-shower observations can probe  sterile neutrinos with suppressed interactions. The Earth-skimming and upward-going channels provide complementary information, with the former determining most of the current flux sensitivity and the latter probing substantially longer Earth chords, establishing eventually physics beyond the Standard Model.

\begin{acknowledgments}

\begin{sloppypar}
The successful installation, commissioning, and operation of the Pierre
Auger Observatory would not have been possible without the strong
commitment and effort from the technical and administrative staff in
Malarg\"ue. We are very grateful to the following agencies and
organizations for financial support:
\end{sloppypar}

\begin{sloppypar}
Argentina -- Comisi\'on Nacional de Energ\'\i{}a At\'omica; Agencia Nacional de
Promoci\'on Cient\'\i{}fica y Tecnol\'ogica (ANPCyT); Consejo Nacional de
Investigaciones Cient\'\i{}ficas y T\'ecnicas (CONICET); Gobierno de la
Provincia de Mendoza; Municipalidad de Malarg\"ue; NDM Holdings and Valle
Las Le\~nas; in gratitude for their continuing cooperation over land
access; Australia -- the Australian Research Council; Belgium -- Fonds
de la Recherche Scientifique (FNRS); Research Foundation Flanders (FWO),
Marie Curie Action of the European Union Grant No.~101107047; Brazil --
Minist\'erio da Ci\^encia, Tecnologia e Inova\c{c}\~ao (MCTI); Czech Republic --
GACR 24-13049S, CAS LQ100102401, MEYS LM2023032,
CZ.02.1.01/0.0/0.0/16{\textunderscore}013/0001402, CZ.02.1.01/0.0/0.0/18{\textunderscore}046/0016010
and CZ.02.1.01/0.0/0.0/17{\textunderscore}049/0008422 and
CZ.02.01.01/00/22{\textunderscore}008/0004632; France -- Centre de Calcul IN2P3/CNRS;
Centre National de la Recherche Scientifique (CNRS); Institut National
de Physique Nucl\'eaire et de Physique des Particules (IN2P3/CNRS);
Germany -- Bundesministerium f\"ur Forschung, Technologie und Raumfahrt
(BMFTR); Deutsche Forschungsgemeinschaft (DFG); Ministerium f\"ur Finanzen
Baden-W\"urttemberg; Helmholtz Alliance for Astroparticle Physics (HAP);
Hermann von Helmholtz-Gemeinschaft Deutscher Forschungszentren e.V.;
Ministerium f\"ur Kultur und Wissenschaft des Landes Nordrhein-Westfalen;
Ministerium f\"ur Wissenschaft, Forschung und Kunst des Landes
Baden-W\"urttemberg; Italy -- Istituto Nazionale di Fisica Nucleare
(INFN); Istituto Nazionale di Astrofisica (INAF); Ministero
dell'Universit\`a e della Ricerca (MUR); CETEMPS Center of Excellence;
Ministero degli Affari Esteri (MAE), ICSC Centro Nazionale di Ricerca in
High Performance Computing, Big Data and Quantum Computing, funded by
European Union NextGenerationEU, reference code CN{\textunderscore}00000013; M\'exico --
Consejo Nacional de Ciencia y Tecnolog\'\i{}a (CONACYT-SECHTI)
No.~CB-A1-S-46703, Universidad Nacional Aut\'onoma de M\'exico (UNAM)
PAPIIT-IN114924; Benem\'erita Universidad Aut\'onoma de Puebla (BUAP), VIEP
and Laboratorio Nacional de Superc\'omputo del Sureste de M\'exico (LNS);
and Benem\'erita Universidad Aut\'onoma de Chiapas (UNACH); The Netherlands
-- Ministry of Education, Culture and Science; Netherlands Organisation
for Scientific Research (NWO); Dutch national e-infrastructure with the
support of SURF Cooperative; Poland -- Ministry of Science and Higher
Education, grant No.~2026/WK/05; National Science Centre, grant
No.~2022/45/B/ST9/02163; Portugal -- Portuguese national funds and FEDER
funds within Programa Operacional Factores de Competitividade through
Funda\c{c}\~ao para a Ci\^encia e a Tecnologia (COMPETE); Romania -- Ministry of
Education and Research, contract no.~30N/2023 under Romanian National
Core Program LAPLAS VII, and grant no.~PN 23 21 01 02; Slovenia --
Slovenian Research and Innovation Agency, grants P1-0031, I0-0033; Spain
-- Ministerio de Ciencia, Innovaci\'on y Universidades/Agencia Estatal de
Investigaci\'on MICIU/AEI /10.13039/501100011033 (PID2022-140510NB-I00,
PCI2023-145952-2, CNS2024-154676, and Mar\'\i{}a de Maeztu CEX2023-001318-M),
Xunta de Galicia (CIGUS Network of Research Centers, Consolidaci\'on
ED431C-2025/11 and ED431F-2022/15) and European Union ERDF; USA --
Department of Energy, Contracts No.~DE-AC02-07CH11359,
No.~DE-FR02-04ER41300, No.~DE-FG02-99ER41107 and No.~DE-SC0011689;
National Science Foundation, Grant No.~0450696, and NSF-2013199; The
Grainger Foundation; Astrophysics Centre for Multi-messenger studies in
Europe (ACME) EU Grant No 101131928; and UNESCO.
\end{sloppypar}

\end{acknowledgments}

\appendix

\section{Energy spectrum of sterile neutrinos}
\label{app:sterile_spectrum}

In this appendix, we derive the energy spectrum of the sterile neutrinos produced in the decay of the dark matter particle $\chi$ induced by the effective interaction in Eq.~\eqref{eqn:Leff}. We consider the decay
\begin{equation}
\chi(P)\to
N_1(p_1)N_1(p_2)
N_2(p_3)N_2(p_4)
\cdots
N_k(p_{2k-1})N_k(p_{2k}) ,
\label{eqn:app_decay}
\end{equation}
and neglect the masses of all final-state sterile neutrinos throughout the calculation. Since there are two $N_1$ particles in the final state, their energy distribution can be obtained from the differential decay width according to
\begin{equation}
\frac{\dif n_{N_1}}{\dif E}
=
\frac{2}{\Gamma}
\frac{\dif\Gamma}{\dif E},
\label{eqn:def_dndE}
\end{equation}
where
\begin{equation}
\dif\Gamma
=
\frac{1}{2M_\chi}
\overline{|\mathcal M|^2}\,
\dif\Phi_{2k}
\end{equation}
is the differential decay width, with $\overline{|\mathcal M|^2}$ denoting the spin-averaged squared matrix element and $\dif\Phi_{2k}$ the Lorentz-invariant $2k$-body phase-space element. Since only the normalized energy spectrum is of interest here, overall normalization factors that are independent of the kinematic variables will be omitted at intermediate steps and restored at the end by imposing
\begin{equation}
\int_0^{M_\chi/2}
\dif E\,
\frac{\dif n_{N_1}}{\dif E}
=2 .
\label{eqn:app_norm_N1}
\end{equation}

\textit{Squared matrix element -- } Up to an overall combinatorial factor associated with the identical particles within each pair, the amplitude can be written as
\begin{equation}
\mathcal{M}
\propto
\prod_{i=1}^{k}
\left[
\overline{u_i^c}(p_{2i-1})
u_i(p_{2i})
\right],
\label{eqn:app_amplitude}
\end{equation}
where $u_i(p)$ denotes the four-component spinor wave function of the fermion $N_i$ with four-momentum $p$, and $u_i^c(p)$ is the corresponding charge-conjugated spinor. Since the $N_i$ are distinct species, the spin sums factorize pair by pair. In the massless limit, this gives 
\begin{equation}
\overline{|\mathcal{M}|^2}
\propto
\prod_{i=1}^{k}
s_i ,
\label{eqn:app_Msq}
\end{equation}
where $s_i\equiv
(p_{2i-1}+p_{2i})^2$ is the invariant mass squared of the $i$th neutrino pair. \\

\textit{Recursive phase-space integration -- } We first integrate over the $2k-2$ particles other than the observed $N_1$. To this end, we introduce the two intermediate four-momenta $S\equiv p_1+p_2$ and $R\equiv\sum_{i=3}^{2k}p_i$ with invariant masses $s\equiv S^2$, $r\equiv R^2$ and $P^2=M_\chi^2$. The $2k$-body phase space can then be factorized using the standard recursive decomposition of multiparticle phase space~\cite{Byckling:1971vca}:
\begin{equation}
\begin{split}
\dif\Phi_{2k}(P)
&=
\frac{\dif s}{2\pi}
\frac{\dif r}{2\pi}
\dif\Phi_2(P;S,R)
\dif\Phi_2(S;p_1,p_2)
\dif\Phi_{2k-2}(R).
\end{split}
\label{eqn:app_recursive_phase_space}
\end{equation}
This decomposition can be interpreted as the sequential decay $\chi(P)\to \chi_1(S)+\chi_2(R)$, followed by $\chi_1(S)\to N_1(p_1)+N_1(p_2)$ and $\chi_2(R)\to N_2(p_3)N_2(p_4)\cdots N_k(p_{2k-1})N_k(p_{2k})$, where $\chi_1$ and $\chi_2$ are pseudo-particles with invariant masses $\sqrt{s}$ and $\sqrt{r}$, respectively.\\

\textit{Two-body phase space --} For completeness, we derive the two-body phase-space element explicitly. We work in the rest frame of the parent particle,
\begin{equation}
P^\mu=(\sqrt{P^2},\mathbf{0}).
\end{equation}
Starting from the generic definition
\begin{equation}
\dif\Phi_2(P;p_a,p_b)
=
(2\pi)^4
\delta^{(4)}(P-p_a-p_b)
\frac{\dif^3\mathbf p_a}{(2\pi)^3 2E_a}
\frac{\dif^3\mathbf p_b}{(2\pi)^3 2E_b},
\label{eqn:app_dPhi2_def}
\end{equation}
we obtain, after integration over $\mathbf{p}_b$,
\begin{equation}
\dif\Phi_2=\frac{1}{16\pi^2}\frac{p_a^2\dif p_a}{E_aE_b}\delta(\sqrt{P^2}-E_a-E_b)\,\dif\Omega .
\label{eqn:app_dPhi2_radial}
\end{equation}
The delta function fixes the magnitude of the three-momentum to
\begin{equation}
p_a=\frac{\lambda^{1/2}(P^2,m_a^2,m_b^2)}{2\sqrt{P^2}},
\label{eqn:app_pa}
\end{equation}
where $\lambda(x,y,z)=x^2+y^2+z^2-2xy-2xz-2yz$ is the K\"all\'en function. Performing the remaining radial and angular integrations then gives
\begin{equation}
\int\dif\Phi_2(P;p_a,p_b)=\frac{\lambda^{1/2}(P^2,m_a^2,m_b^2)}{8\pi P^2} .
\label{eqn:app_Phi2_general}
\end{equation}
Therefore, for the first two-body decay $\chi(P)\to\chi_1(S)+\chi_2(R)$, 
we  have
\begin{equation}
\int\dif\Phi_2(P;S,R)=\frac{\lambda^{1/2}(P^2,s,r)}{8\pi P^2} .
\label{eqn:app_Phi2_SR}
\end{equation}
For the subsequent decay $\chi_1(S)\to N_1(p_1)+N_1(p_2)$,
both final-state particles are massless. The delta function in the
two-body phase space therefore fixes
\begin{equation}
E_1^\star=E_2^\star=\frac{\sqrt{s}}{2},
\qquad
|\mathbf p_1^\star|
=
|\mathbf p_2^\star|
=
\frac{\sqrt{s}}{2},
\label{eqn:app_pair_energies}
\end{equation}
in the $\chi_1$ rest frame. Since $\lambda^{1/2}(s,0,0)=s$, we obtain 
\begin{equation}
\label{eqn:app_dPhi2_pair_cos}
\dif\Phi_2(S;p_1,p_2)=\frac{1}{16\pi}\dif\cos\theta^\star
\end{equation}
The polar angle $\theta^\star$ will be related to the energy of $N_1$
in the $\chi$ rest frame below.\\

\textit{Integration over the unobserved phase space --} It is useful to define at this stage the weighted phase-space integral
\begin{equation}
{\cal W}_m(P^2)
\equiv
\int\dif\Phi_{2m}(P)
\prod_{i=1}^{m}s_i ,
\label{eqn:app_Wdef}
\end{equation}
where the $2m$ final-state particles are grouped into $m$ pairs. Once all final-state momenta have been integrated over, ${\cal W}_m$ is a Lorentz scalar and therefore depends only on the invariant mass $P^2$. The recursive phase-space decomposition leads to
\begin{equation}
{\cal W}_m(P^2)=\frac{1}{(2\pi)^2}\int \dif s\,\dif r \frac{\lambda^{1/2}(P^2;s,r)} {8\pi P^2} \frac{1}{8\pi} s{\cal W}_{m-1}(r),
\label{eqn:app_Wrecursion}
\end{equation}
where the integration region is determined by the condition that the two intermediate states can be produced, $s\geq 0$, $r\geq 0$ and $\sqrt{s}+\sqrt{r}\leq\sqrt{P^2}$. 

The dependence of ${\cal W}_m$ on the total invariant mass can be
established directly from this recursion. For $m=1$,
\begin{equation}
{\cal W}_1(P^2)=\int\dif\Phi_2(P;p_1,p_2)s_1=\frac{P^2}{8\pi}.
\label{eqn:app_W1}
\end{equation}
Assuming that
\begin{equation}
{\cal W}_{m-1}(r)=A_{m-1}r^{3m-5},
\label{eqn:app_Winduction}
\end{equation}
where $A_{m-1}$ is independent of $r$, and introducing the dimensionless variables $u=s/P^2$ and $v=r/P^2$, we obtain the recursive expression   
\begin{equation}
\begin{split}
{\cal W}_m(P^2)
&=
\frac{A_{m-1}}
{256\pi^4}
(P^2)^{3m-2}
\int\dif u\,\dif v\,
uv^{3m-5}
\lambda^{1/2}(1,u,v).
\end{split}
\label{eqn:app_Wdimensionless}
\end{equation}
The remaining integral is dimensionless and depends only on $m$.
Consequently,
\begin{equation}
{\cal W}_m(P^2)=A_m(P^2)^{3m-2},
\label{eqn:app_Wscaling}
\end{equation}
which is the dependency on $P^2$ sought for. For the $2k-2$ unobserved particles in Eq.~\eqref{eqn:app_decay}, this
result gives
\begin{equation}
\int\dif\Phi_{2k-2}(R)\prod_{i=2}^{k}s_i=A_{k-1}r^{3k-5}.
\label{eqn:app_unobserved_integral}
\end{equation}
Using this expression, the differential decay
width can therefore be written, up to an overall normalization, as
\begin{equation}
\dif\Gamma
\propto
sr^{3k-5}
\frac{\dif s\,\dif r}{(2\pi)^2}
\dif\Phi_2(P;S,R)
\dif\Phi_2(S;p_1,p_2).
\label{eqn:app_reduced_rate}
\end{equation}

\textit{Energy distribution of $N_1$ --} We finally derive the distribution in the energy $E_1$ of one of the
$N_1$ particles. We work in the $\chi$ rest frame, $P^\mu=(M_\chi,\mathbf{0})$.
For the decay $\chi(P)\to\chi_1(S)+\chi_2(R)$,
energy-momentum conservation gives
\begin{equation}
E_S=\frac{M_\chi^2+s-r}{2M_\chi},
\qquad
|\mathbf S|=\frac{\lambda^{1/2}(M_\chi^2,s,r)}{2M_\chi}.
\label{eqn:app_S_kinematics}
\end{equation}
In the rest frame of $\chi_1(S)$, Eq.~\eqref{eqn:app_pair_energies}
leads to
\begin{equation}
E_1^\star=|\mathbf p_1^\star|=\frac{\sqrt{s}}{2}.
\end{equation}
The boost from the $\chi_1$ rest frame to the $\chi$ rest frame has relativistic factors $\gamma_S=E_S/\sqrt{s}$ and $\beta_S=\left|\mathbf S\right|/E_S$. Consequently,
\begin{equation}
E_1=\frac{E_S}{2}+\frac{|\mathbf S|}{2}\cos\theta^\star .
\label{eqn:app_Eboost}
\end{equation}
It is convenient at this stage to introduce the dimensionless variables $x\equiv 2E_1/M_\chi$, $a\equiv s/M_\chi^2$ and $b\equiv r/M_\chi^2$.
Eq.~\eqref{eqn:app_Eboost} then becomes
\begin{equation}
x=\frac{1+a-b}{2}+\frac{\lambda^{1/2}(1,a,b)}{2}\cos\theta^\star .
\label{eqn:app_xcostheta}
\end{equation}
To obtain the distribution in $x$, we insert the identity
\begin{equation}
1=\int\dif x\,\delta\left[x-\frac{1+a-b}{2}-\frac{\lambda^{1/2}(1,a,b)}{2}\cos\theta^\star
\right]
\end{equation}
into the angular integration in Eq.~\eqref{eqn:app_dPhi2_pair_cos}. The integration over $\cos\theta^\star$ yields
\begin{multline}
\int\dif\cos\theta^\star \,\frac{\dif\Phi_2(S;p_1,p_2)}{\dif\cos\theta^\star}=\\
\frac{1}{8\pi\lambda^{1/2}(1,a,b)}\Theta\left[1-\left|\frac{2x-1-a+b}{\lambda^{1/2}(1,a,b)}\right| \right]\dif x,
\label{eqn:app_dPhi2_x}
\end{multline}
with $\Theta$ the Heaviside function. The factor $\lambda^{-1/2}$ in Eq.~\eqref{eqn:app_dPhi2_x} is cancelled by the factor $\lambda^{1/2}$ in $\dif\Phi_2(P;S,R)$ (Eq.~\eqref{eqn:app_Phi2_SR}). The remaining kinematic dependence is therefore contained in the matrix-element factor and in the boundaries of the $(s,r)$ integration.

The condition imposed by the step function is
\begin{equation}
\left|
2x-1-a+b
\right|
\leq
\lambda^{1/2}(1,a,b).
\label{eqn:app_physical_condition}
\end{equation}
For $0\leq x\leq1$, this condition is equivalent to
\begin{equation}
0\leq a\leq x,
\qquad
0\leq b\leq
\frac{(1-x)(x-a)}{x}.
\label{eqn:app_ab_limits}
\end{equation}
Combining the matrix-element factor $s$ associated with the observed
$N_1N_1$ pair with the result of the integration over the remaining
particles, the energy distribution therefore reduces to
\begin{equation}
\frac{\dif\Gamma}{\dif x}\propto\int_0^x\dif a\,a \int_0^{(1-x)(x-a)/x}\dif b\,b^{3k-5},
\label{eqn:app_final_integral}
\end{equation}
which yields to the spectrum 
\begin{equation}
\frac{1}{\Gamma}\frac{\dif\Gamma}{\dif x}=\frac{x^2(1-x)^{3k-4}}{B(3,3k-3)},  
\label{eqn:app_single_spectrum}
\end{equation}
with $B(x,y)$ the Euler beta function. Finally, given that  
\begin{equation}
\frac{\dif n_{N_1}}{\dif E}=\frac{2}{\Gamma}\frac{\dif\Gamma}{\dif E}=\frac{4}{M_\chi}\frac{1}{\Gamma}\frac{\dif\Gamma}{\dif x},
\end{equation}
we obtain  
\begin{equation}
\frac{\dif n_{N_1}}{\dif E}=\frac{4}{B(k,3k-3)M_\chi}\left(\frac{2E}{M_\chi}\right)^2\left(1-\frac{2E}{M_\chi}\right)^{3k-4},
\label{eqn:app_final_spectrum}
\end{equation}
which is Eq.~\eqref{eqn:dnNdE}.




\bibliography{apssamp}

\end{document}